\documentclass{LMCS}

\def\dOi{10(1:12)2014}
\lmcsheading%
{\dOi}
{1--29}
{}
{}
{Sep.~30, 2012}
{Feb.~13, 2014}
{}

\ACMCCS{[{\bf Theory of computation}]: Logic, Formal languages and
  automata theory; [{\bf Software and its engineering}]: Software
  organization and properties---Software functional
  properties---Formal methods; [{\bf Hardware}]: Hardware
  validation---Functional verification---Theorem proving and SAT
  solving}

\usepackage{amsmath,amssymb,amsfonts}
\usepackage{graphicx}
\usepackage{pstricks,pst-node}
\usepackage{temporal}
\usepackage{xspace}
\usepackage{caption}
\usepackage{subcaption}
\usepackage{tikz}
\usepackage{hyperref}
  \usetikzlibrary{automata,positioning}
\newcommand{\Buchi}{B\"uchi\xspace}

\newcommand{\bigdotcup}{\ensuremath{\mathaccent\cdot{\bigcup}}}

\newcommand{\token}{\mathsf{tok}}
\newcommand{\send}{\mathsf{send}}
\newcommand{\sched}{\mathsf{sched}}
\newcommand{\spec}{\varphi}
\newcommand{\pspec}{\Phi}
\newcommand{\mU}{\mathcal{U}}

\newcommand{\impl}{\, \rightarrow \, }
\newcommand{\card}[1]{{|}#1{|}}
\newcommand{\bbN}{\mathbb{N}}
\newcommand{\bbB}{\mathbb{B}}
\newcommand{\mT}{\mathcal{T}}
\newcommand{\mG}{\mathcal{G}}

\begin{document}

\title{Parameterized Synthesis}
\author[S.~Jacobs]{Swen Jacobs}
\address{IAIK, Graz University of Technology, Austria}
\email{\{swen.jacobs,roderick.bloem\}@iaik.tugraz.at}
\thanks{This work was supported in part by the European Commission through project 
		DIAMOND (FP7-2009-IST-4-248613), and by the
    Austrian Science Fund (FWF) under the RiSE National Research
    Network (S11406).}

\author[R.~Bloem]{Roderick Bloem}
\address{\vspace{-18 pt}}


\begin{abstract}
  We study the synthesis problem for distributed architectures with a
  parametric number of finite-state components.  
	Parameterized
  specifications arise naturally in a synthesis setting, but thus far
  it was unclear how to detect realizability and how to perform
  synthesis in a parameterized setting. 
	Using a classical result from verification, we show that
  for a class of specifications in indexed LTL\textbackslash X, parameterized synthesis
  in token ring networks is equivalent to distributed synthesis in a
  network consisting of a few copies of a single process.  Adapting a
  well-known result from distributed synthesis, we show that the latter
  problem is undecidable. We describe a semi-decision procedure for
  the parameterized synthesis problem in token rings,
  based on bounded synthesis. We extend the approach to parameterized
  synthesis in token-passing networks with arbitrary topologies, and
  show applicability on a simple case study. Finally, we sketch a
  general framework for parameterized synthesis based on cutoffs and
  other parameterized verification techniques. 
\end{abstract}

\maketitle

\section{Introduction}
\label{sec:intro}

Synthesis is the problem of turning a temporal logical specification into a
reactive system \cite{Church62,Pnueli89}.  In synthesis, parameterized
specifications occur very naturally.  For instance, Piterman, Pnueli,
and Sa'ar~\cite{Piterm06b} illustrate their GR(1) approach with two parameterized
examples of an arbiter and an elevator controller.
Similarly, the case studies considered by Bloem et
al.~\cite{Bloem07,Bloem07b} consist 
of a parameterized specification of the AMBA bus arbiter
and a parameterized generalized buffer. 
A simple example of a parameterized specification is
\medskip
\[
\begin{array}{lll}
& \forall i. & \always(r_i \rightarrow \eventually g_i)\\
\wedge & \forall i \neq j. & \always(\neg g_i \vee \neg g_j).
\end{array}
\]

\medskip
\noindent
This specification describes an arbiter serving an arbitrary number of
clients, say $n$.  Client $i$ controls a signal $r_i$ for sending requests and
can read a signal $g_i$ for receiving grants.  The specification states that, for each
client $i$, each request $r_i$ is eventually followed by a grant $g_i$,
but grants never occur simultaneously.

Most previous approaches have focused on the synthesis of such systems for
a fixed $n$.  The question whether such a specification is realizable
for \emph{any} $n$ is natural: it occurs, for instance, in the work
on synthesis of processes for the leader election problem by Katz and
Peled~\cite{Katz09}. Only an answer to this question
can determine whether a parameterized specification is correct. A
further natural question is how to construct a parameterized system,
i.e., a recipe for quickly constructing a system for an arbitrary
$n$. Such a construction would avoid the steep increase of runtime and
memory use with $n$ that current tools incur
\cite{Bloem07,Bloem07b,Finkbeiner12}. 

Parameterized systems have been studied extensively in the context of
verification.  It is well known that the verification of such systems
is in general undecidable \cite{Apt86,Suzuki88}, but several decidable cases
have been identified. In particular, for restricted topologies like 
token-passing networks, the
problem of verifying a network of isomorphic processes of arbitrary
size can be reduced to the verification of a fixed, small network. As a corollary, 
synthesis in a network with an arbitrary number of processes can be
reduced to synthesis in a small network, as long as the restricted
topology is respected. In this paper, we focus first on token rings
\cite{Emerso03}, and consider general token-passing networks~\cite{Clarke04c} later.

For token rings, the parameterized synthesis problem is equivalent to
distributed synthesis in a small network of isomorphic processes. This
question is closely related to that of distributed synthesis
\cite{Pnueli90,Finkbeiner05,Schewe06}.  Distributed synthesis is undecidable for
all systems in which processes are incomparable with respect to their information about the
environment.
Our problem is slightly different in that we 
only consider specifications in LTL\textbackslash X and that our
synthesis problem is \emph{isomorphic}, i.e., processes have to be
identical.  Unfortunately, this problem, and thus the original problem
of parameterized synthesis, is also undecidable.


Having obtained a negative decidability result, we turn our attention
to a semi-decision procedure, namely the bounded synthesis approach of Finkbeiner and 
Schewe~\cite{Finkbeiner12a}. The bounded synthesis method searches for systems
with a bounded number of states.  We modify this approach to deal with
isomorphic token-passing systems.  Bounded synthesis reduces the
problem of realizability to an SMT formula, a model of which gives an
implementation of the system. 

As a generalization of token-rings, we consider token-passing networks
with arbitrary topologies~\cite{Clarke04c}, and
show how to extend verification results to synthesis in these
networks. Also, we propose a symmetry reduction technique suitable for
both verification and synthesis in these networks.

As a proof of concept, we use an SMT solver to
synthesize a simple parameterized arbiter in both a token ring and a
more general token-passing network. We show that a minimal
implementation can be synthesized in reasonable time. 

Finally, we argue that our approach is not limited to token-passing
networks, but can be seen as a framework to lift other 
classes of systems and specifications, in particular those that allow a cutoff
for the corresponding verification problem.

\subsubsection*{Related Work.}
There have been previous approaches to solve parameterized synthesis problems~\cite{Emerson90,Attie98}. However, among other restrictions, these results only consider cases where information about the environment is
  essentially the same for all processes. Therefore, the resulting
  synthesis problems are decidable for a fixed or even an arbitrary
  number of processes.

\section{Preliminaries}
\label{sec:prelim}

We assume that the reader is familiar with LTL, the synthesis problem,
and the basic idea of parameterized model checking. For understanding
the technical details of our approach, knowledge about the
\emph{bounded synthesis} method~\cite{Finkbeiner12a} is helpful. Also, we build on the decidability results for parameterized token rings of Emerson and Namjoshi~\cite{Emerso03}, as well as those for parameterized token networks by Clarke et al.~\cite{Clarke04c}.


\subsection{Distributed Reactive Systems}
\label{sec:prelim-distributed}

\subsubsection*{Architectures} An \emph{architecture} $A$ is a tuple
$(P,env,V,I,O)$, where $P$ is a finite set of 
processes, containing the environment process $env$ and system processes $P^- =
P \setminus \{env\}$, $V$ is a set of 
Boolean system variables, $I = \left\{I_i 
\subseteq V ~|~ i \in P^- \right\}$ assigns a set $I_i$ of Boolean input variables to each
system process, and $O = \left\{ O_i \subseteq V ~|~ i \in P \right\}$ assigns a
set $O_i$ of Boolean output variables to each process, such that $\bigdotcup_{i \in P}
  O_i = V$. In contrast to output variables, inputs may be shared
between processes. Without loss of generality, we use natural numbers to refer to
system processes, and assume $P^-=\{1,\ldots,k\}$ for an architecture with $k$
system processes. We denote by $\mathcal{A}$ the set of all architectures.

\subsubsection*{Implementations.}
An \emph{implementation} $\mT_i$ of a system process $i$
  with inputs $I_i$ and outputs $O_i$ is a labeled transition system
  (LTS) $\mT_i = (T_i, t_i, \delta_i, o_i)$, where $T_i$ is a set of states including the
  initial state $t_i$, $\delta_i: T_i \times \mathcal{P}(I_i) \rightarrow T_i$ a transition function,
  and $o_i: T_i \rightarrow \mathcal{P}(O_i)$ a labeling
  function. $\mT_i$ is a \emph{finite} LTS if $T_i$ is finite.

The \emph{composition} of the 
  set of system process implementations $\left\{ \mT_1, \ldots, \mT_k \right\}$
  is the LTS $\mT_A = (T_A,t_0,\delta,o)$, where the states are $T_A = T_1 \times
  \dots \times T_k$, the initial state $t_0 = (t_1, \ldots, t_k)$,
  the labeling function $o : T_A \rightarrow \mathcal{P}(\bigdotcup_{1\leq i \leq k} O_i)$
  with $o(t_1,\ldots,t_k) = o_1(t_1) \cup \dots \cup o_k(t_k)$, and finally the
  transition function $\delta : T_A \times \mathcal{P}(O_{env}) \rightarrow T_A$ with 
\[ \delta((t_1,\ldots,t_k),e) = (\delta_1(t_1,(o(t_1,\ldots,t_k) \cup e) \cap I_1),
  \ldots,  \delta_k(t_k,(o(t_1,\ldots,t_k) \cup e) \cap I_k)).\]
  That is, every
  process advances according to its own transition function and input variables,
  where inputs from other system processes are interpreted according to the
  labeling of the current state.

A \emph{run} of an LTS $\mT = (T,t_0,\delta,o)$ is an infinite sequence
$(t^0,e^0),(t^1,e^1),\ldots$, where $t^0 = t_0$, $e^i \subseteq
O_{env}$ and $t^{i+1} = \delta(t^i,e^i)$.  $\mT$ \emph{satisfies} an LTL
formula $\spec$ if for every run of $\mT$, the sequence $o(t^0) \cup e^0,
o(t^1)\cup e^1,\dots$ is a model of $\spec$.


\subsubsection*{Asynchronous Systems.} An \emph{asynchronous system} is an LTS such that
in every transition, only a subset of the system processes changes their
state. This is decided by a \emph{scheduler} that can choose for every
transition which of the processes (including the environment) is allowed to make a step. In
our setting, we will assume that the environment is always scheduled, and
consider the scheduler as a part of the environment.

Formally, $O_{env}$ contains additional scheduling
variables $s_1, \ldots, s_k$, and $s_i \in I_i$ for every $i$. For any
$i \in P^-,t \in T_i$ and $I \subseteq I_i$, we require $\delta_i(t,I) = t$ whenever $s_i \not \in I$.

\subsubsection*{Token Rings.} We consider a class of architectures called
\emph{token rings}, where the only communication between system processes is a
token. At any time only one 
process can possess the token, and a process $i$ that has the token can decide to pass
it to process $i+1$ by raising an output $\send_i \in O_i \cap 
I_{i+1}$. For processes in token rings of size $k$, addition and
subtraction is done modulo $k$.

 We assume that token rings are implemented as
asynchronous systems, where in every step only one system process may change its state,
except for token-passing steps, in which both of the involved processes change
their state. 

\subsection{Synthesis Problems}
\subsubsection*{Distributed Synthesis.}
The \emph{distributed synthesis problem} for a given architecture $A$ and a
specification $\spec$ is to find implementations for the system processes of
$A$, such that the composition of the implementations $\mT_1,\ldots,\mT_k$ satisfies
$\spec$, written $A,(\mT_1,\ldots,\mT_k) \models \spec$.  A
specification $\spec$ is \emph{realizable} with respect to an architecture $A$
if such implementations exist. 

Checking realizability of LTL
specifications has been shown to be undecidable for architectures in
which processes have incomparable information about the environment 
in the synchronous case~\cite{Finkbeiner05}, and even for
all architectures with more than one system process in the
asynchronous case~\cite{Schewe06}.

\subsubsection*{Bounded Synthesis.} 
The \emph{bounded synthesis problem} for given architecture $A$, specification
$\spec$ and a set of bounds $\{b_i \in \bbN ~|~ i \in P^- \}$ on the size of
system processes as well as a bound $b_A$ for the composition $\mT_A$, is to find
implementations $\mT_i$ for the system processes such that their composition
$\mT_A$ satisfies $\spec$, with $|T_i| \leq b_i$ for all process
implementations, and $|T_A| \leq b_A$.

\section{Parameterized Synthesis}
\label{sec:synthesis}

In this section, we introduce the parameterized synthesis problem.  Using a
classical result for the verification of token rings by Emerson and
Namjoshi~\cite{Emerso03}, we show that parameterized synthesis for token
ring architectures and specifications in LTL\textbackslash X can be reduced to
distributed synthesis of isomorphic processes in a ring of fixed size. We then
show that, for this class 
of architectures and specifications, the isomorphic distributed synthesis
problem is still undecidable.

\newpage
\subsection{Definition}
\label{sec:synthesis:definition}
%
%
\subsubsection*{Parameterized Architectures and Specifications.} 
 A \emph{parameterized
  architecture} is a function $\Pi: \bbN \rightarrow \mathcal{A}$. A
\emph{parameterized token ring} is a parameterized architecture $\Pi_R$
with $\Pi_R(n) =(P_n,env,V_n,I_n,O_n)$, where 

\begin{itemize}
\item $P_n = \{ env, 1, \ldots, n \}$,
\item $I_n$ is such that all system processes are assigned isomorphic sets of
inputs, consisting of the token-passing input $\send_{i-1}$ 
and a set of inputs from the environment, distinguished by indexing each input
with $i$.
\item Similarly, $O_n$ assigns isomorphic, indexed sets of outputs to all system
  processes, with $\send_i \in O_n(i)$, and
every output of $env$ is indexed with all values from $1$ to $n$.
\end{itemize}
%
%
A \emph{parameterized specification} $\pspec$ is a sentence in indexed
LTL, that is, an LTL specification
with indexed variables and a combination of universal and existential
quantification (in prenex form) over all indices. We say that
a parameterized architecture $\Pi$ and a process implementation $\mT$
\emph{satisfy} a parameterized specification (written $\Pi,\mT \models \pspec$)
if for all $n$, $\Pi(n),(\mT,\ldots,\mT) \models \pspec$.

\begin{exa}
Consider the parameterized token ring $\Pi_{arb}$ with $\Pi_{arb}(n) =$\\ $(P_n,env,V_n,I_n,O_n)$, where

\begin{align}
P_n &=  \{ env, 1, \ldots, n \}\\
V_n &=  \{ r_1, \ldots, r_n, g_1 \ldots, g_n, \send_1, \ldots, \send_n \}\\
I_n(i) &=  \{ r_i, \send_{i-1} \}\\
O_n(env) &=  \{ r_1, \ldots, r_n \}\\
O_n(i) &=  \{ g_i, \send_i \}
\end{align}

The architecture $\Pi_{arb}(n)$ defines a token ring with $n$ system
processes, with each process $i$ receiving an input $r_i$ from the environment
and another input $\send_{i-1}$ from the previous process in the ring, and an
output $\send_i$ to the next process, as well as an output $g_i$ to the environment.

An instance of this parameterized architecture for $n=4$ is depicted in
Figure~\ref{fig:ring-architecture}. Together with the parameterized specification
from Section~\ref{sec:intro}, we will use it in
Section~\ref{sec:implementation} to synthesize process implementations for a
parameterized arbiter.
\end{exa}

\begin{figure}
\begin{pspicture}(0,-1)(4,4.5)
\psset{arrows=->,nodesep=5pt}
\put(1.5,5){\rnode{a1}{}}
\cnodeput(1.5,3.5){1}{$1$}
\put(1.5,2){\rnode{b1}{}}
\cnodeput(4,2){2}{$2$}
\put(4,3.5){\rnode{a2}{}}
\put(4,0.5){\rnode{b2}{}}
\cnodeput(2,0){3}{$3$}
\put(2,1.5){\rnode{a3}{}}
\put(2,-1.5){\rnode{b3}{}}
\cnodeput(-0.5,1.5){4}{$4$}
\put(-0.5,3){\rnode{a4}{}}
\put(-0.5,0){\rnode{b4}{}}

\ncline{a1}{1}\tlput{$r_1$}
\ncline{1}{b1}\tlput{$g_1$}
\ncline{1}{2}\taput{$\send_1$}
\ncline{a2}{2}\trput{$r_2$}
\ncline{2}{b2}\trput{$g_2$}
\ncline{2}{3}\taput{$\send_2$}
\ncline{a3}{3}\tlput{$r_3$}
\ncline{3}{b3}\tlput{$g_3$}
\ncline{3}{4}\taput{$\send_3$}
\ncline{a4}{4}\tlput{$r_4$}
\ncline{4}{b4}\tlput{$g_4$}
\ncline{4}{1}\taput{$\send_4$}

\end{pspicture}
\caption{Token ring architecture with 4 processes}
\label{fig:ring-architecture}
\end{figure}

\subsubsection*{Isomorphic and Parameterized Synthesis.}
The \emph{isomorphic synthesis problem} for an architecture $A$ and a 
specification $\spec$ is to find an implementation $\mT$ for all system
processes $(1,\ldots,k)$ such that $A,(\mT,\ldots,\mT) \models \spec$, also 
written $A, \mT \models \spec$.
The \emph{parameterized synthesis problem} for a parameterized
architecture $\Pi$ and a parameterized specification $\pspec$ is
to find an implementation $\mT$ for all system processes such
that $\Pi,\mT \models \pspec$.  The \emph{parameterized (isomorphic)
  realizability problem} is the question whether such an implementation exists.

\subsection{Reduction of Parameterized to Isomorphic Synthesis}
\label{sec:synthesis:reduction}

Emerson and Namjoshi \cite{Emerso03} have shown that verification of LTL\textbackslash X
properties for implementations of parameterized token rings can be
reduced to verification of a small ring with up to five processes,
depending on the form of the specification.\footnote{Actually, the
  result by Emerson and Namjoshi is for CTL*\textbackslash X, but for
  synthesis we only consider the fragment LTL\textbackslash X.} For a sequence
$\overline{t}$ of index variables and terms in arithmetic modulo $n$,
let $\spec(\overline{t})$ in the following be a formula in
LTL\textbackslash X that only refers
to system variables indexed by terms in $\overline{t}$.

\begin{thm}[\cite{Emerso03}]
\label{thm:reduction}
Let $\Pi_R$ be a parameterized token ring, $\mT$ an
implementation of the isomorphic system processes that ensures fair token
passing, and $\pspec$ a parameterized specification.  Then 
\begin{enumerate}[label=\alph*)]
\item If $\pspec = \forall i.~ \spec(i)$, then 
\[\Pi_R,\mT \models
  \pspec \iff \textrm{ For } 1 \leq k \leq 2:\ \Pi_R(k), \mT \models \bigwedge_{1 \leq i \leq k} \spec(i)
  .\]
\item If $\pspec = \forall i.~ \spec(i,i+1)$, then 
\[ \Pi_R,\mT \models \pspec \iff \textrm{ For } 1 \leq k \leq 3:\ \Pi_R(k),\mT \models \bigwedge_{1 \leq i
  \leq k} \spec(i,i+1).\]
\item If $\pspec = \forall i \neq j.~ \spec(i,j)$, then 
\[ \Pi_R,\mT \models \pspec \iff \textrm{ For } 1 \leq k \leq 4:\ \Pi_R(k),\mT \models
\bigwedge_{\substack{1 \leq i,j \leq k\\ i \neq j}} \spec(i,j).\]
\item If $\pspec = \forall i \neq j.~ \spec(i,i+1,j)$, then 
\[\Pi_R,\mT \models \pspec \iff \textrm{ For } 1 \leq k \leq 5:\ \Pi_R(k),\mT \models
\bigwedge_{\substack{1 \leq i,j \leq k\\ i \neq j}} \spec(i,i+1,j).\]
\end{enumerate}
\end{thm}
\noindent This theorem implies that verification of such structures is
decidable. For synthesis, we obtain the following corollary:

\begin{cor}
\label{cor:reduction}
For a given parameterized token ring $\Pi_R$ and parametric
specification $\pspec$, parameterized synthesis can be reduced to isomorphic
synthesis in rings of size up to $2$ (3, 4, 5) for specifications of type a)
(b, c, d, respectively).
\end{cor}

In the following, we will show that this reduction in general does not make the
synthesis problem decidable.

\subsection{Decidability}
\label{sec:synthesis:decidability}
The parameterized synthesis problem is closely related to the distributed
synthesis problem~\cite{Pnueli90,Finkbeiner05}. We will use a modification of
the original undecidability proof for distributed systems to show undecidability
of isomorphic realizability in token rings, which in turn implies undecidability of
parameterized realizability.

\begin{thm}
\label{thm:undecidable}
The isomorphic realizability problem is undecidable for
token rings with 2 or more processes and specifications in LTL\textbackslash
X. 
\end{thm}

\proof
We first reconsider the undecidability proof for synchronous
distributed processes by Pnueli and Rosner~\cite{Pnueli90}, and then
show how to modify the construction to prove undecidability in our
setting.

\noindent {\bf Standard undecidability proof.} Pnueli and Rosner 
have shown that distributed realizability is undecidable for two
synchronous processes, neither of 
which can observe the inputs or outputs of the other.\footnote{For a
  generalization of the undecidability proof to all architectures with
  ``information forks'' we refer to Finkbeiner and
  Schewe~\cite{Finkbeiner05}.} The 
undecidability proof reduces the halting problem for deterministic
Turing machines to the distributed realizability problem. This is done by
encoding a specification in LTL that forces both processes to each
simulate the given Turing machine $M$, and halt.\footnote{Pnueli and
  Rosner argue informally that such specifications
  can be expressed in LTL. For a more complete treatment and an
  analysis of temporal logic fragments that are sufficient to express
  such specifications, we refer to Schewe~\cite{Schewe2013}.}

For notational simplicity, assume that processes have outputs
sufficient to encode 
configurations of $M$ (i.e., a valuation of the process outputs
represents a ${\sf tape}$ symbol, a ${\sf state}$ symbol, or a ${\sf
  blank}$ symbol). For configurations $C,D$, represented as 
  sequences of these output symbols, denote by $C \vdash D$ that $D$ is a
  legal successor configuration of $C$. Each process $i \in \{1,2\}$
  has a single input ${\sf start}_i$ from the environment. At any
  given point in time, let $L_i$ be the number of ${\sf start}_i$
  signals the environment has sent thus far. Consider the following
  assumptions on the environment inputs:
\begin{enumerate}
\item The environment only sends ${\sf start}_i$ if both processes
  currently send a ${\sf blank}$.
\item At any time, $\card{L_1 - L_2} \leq 1$.
\end{enumerate}

Let $\psi$ denote the conjunction of these environment
assumptions. Then, consider the following specification of the processes:
\begin{enumerate}
\item Process $i$ outputs ${\sf blank}$ symbols until it receives the
  first ${\sf start}_i$ signal.
\item Whenever process $i$ receives a ${\sf start}_i$, in the
  following state it will start to output a legal configuration of
  $M$, followed by ${\sf blank}$ symbols until it receives the next
  ${\sf start}_i$.
\item After receiving the first ${\sf start}_i$, process $i$ outputs
  the initial configuration of $M$.
\item Assume the processes receive ${\sf start}_1$ and ${\sf start}_2$
  at the same time, and denote by $C$ and $D$ the configurations that processes
  $1$ and $2$ start to output now. Then:
  \begin{enumerate}
  \item $C = D$ if $L_1 = L_2$,
  \item $C \vdash D$ if $L_2 = L_1 + 1$, and
  \item $D \vdash C$ if $L_1 = L_2 + 1$.
  \end{enumerate}
\end{enumerate}

The crucial part is the last statement: since the processes cannot
observe in- or outputs of each other (and thus cannot know which one
of them ``goes first'', if any), requirement (a) forces them to
produce the same 
outputs if given the same inputs, and (b) and (c) together force them
to correctly simulate $M$.
Let $\varphi$ denote the conjunction of these statements about the
processes. Then, every system which realizes the specification given
as $\psi \impl \varphi$ must consist of two processes each satisfying
the following (by \cite[Lemma 4.3]{Pnueli90}):
\begin{itemize}
\item The process outputs ${\sf blank}$ symbols until it receives the
  first ${\sf start}$ signal.
\item If the process receives a ${\sf start}$ signal and has received
  $k$ ${\sf start}$ signals before, then the process starts to output
  configuration $C_{k+1}$ of $M$ (where $C_1 \vdash C_2 \vdash \dots$
  is the sequence of configurations of $M$ on the empty input tape).
\end{itemize}

\noindent Thus, to satisfy $\psi \impl \varphi$, each of the two processes must
correctly output the complete run of $M$, with configurations
separated by a number of ${\sf blank}$ symbols. If we add to the
specification that the process must eventually output a ${\sf halt}$
symbol (standing for the halting state of $M$), then the new
specification is realizable if and only if $M$ halts. In
  particular, if $M$ halts, then the specification is finite-state
  realizable, since only finitely many steps need to be simulated.
Thus, this encoding reduces the halting problem of deterministic
Turing machines to the realizability problem of distributed,
synchronous finite-state processes.

\smallskip
\noindent {\bf Modifications for isomorphic realizability of LTL\textbackslash X
  specifications in token rings.}
To prove the statement of Theorem~\ref{thm:undecidable}, we amend the
construction from above such that the halting problem of deterministic
Turing machines is reduced to our modified realizability problem. We
need to consider the following modifications:
\begin{enumerate}
\item The composition of the two processes is asynchronous, and
  the processes can communicate by passing a token.
\item We are not allowed to use the $\nextt$ operator in the specification.
\item We restrict to the (possibly simpler) isomorphic synthesis
  problem.
\end{enumerate}

To handle the first point, we force the system to use the token for
synchronization of processes. That is, one step of the system from the
original proof corresponds to one cycle of the token in the new
system. To ensure that every infinite run also has infinitely many
cycles of the token, we need the usual assumption of fair
scheduling and require fair token passing of the processes.

Additionally, we augment the specification to assume that the token
starts at a designated process, say $1$. Furthermore, we require
\emph{restricted output modification}: each process changes its output
only at the moment it receives the 
token, i.e., only once in each full cycle of the token. For every
possible output symbol ${\sf out}_i$ of process $i$, this can be
expressed as  

\[ {\sf out}_i \leftrightarrow \left({\sf out}_i \weakuntil \left(\neg \token_i \land
({\sf out}_i \weakuntil \token_i)\right)\right).
\]

We also assume that the environment keeps all ${\sf start}_i$ signals
constant during a cycle. That is, we call states where process $1$ has
just received the token \emph{$1$-receiving states}, and assume that
environment inputs only change when entering a $1$-receiving state.

Note that the token cannot be used to pass any additional information
(beyond synchronization): the only freedom a process has is
\emph{when} to pass the token, and by lack of a global clock and
visibility of the output signals of the other processes, a given
process cannot measure this time or observe any changes of the system
during this time.

To handle the loss of the $\nextt$ operator, we use the assumption of
restricted output modification to correlate successive states of the
original synchronous system to successive $1$-receiving states of the
asynchronous system. That is, 
$\nextt \spec$ for the synchronous system corresponds to
\[ \left(\token_1 \land \left( \token_1 \weakuntil \left( \neg \token_1 \land \left(\neg \token_1 \weakuntil (\token_1 \land \spec)\right) \right) \right) \right) \lor \left( \neg \token_1 \land \left( \neg \token_1 \weakuntil (\token_1 \wedge \spec) \right) \right)\]
for the asynchronous system.
We replace all occurrences of the form $\nextt \spec$ in the original
specification by the corresponding instance of the formula
above. Thus, every statement that originally referred to the next state
now refers to the next $1$-receiving state. As by restricted output
modification none of the in- or outputs 
of the system will change between two such states, the rest of the
specification is satisfied by a given run of the asynchronous system
if and only if it is satisfied by the projection of this run
to $1$-receiving states.


In summary, the modified specification forces the processes to
simulate the Turing machine $M$ in the following sense: the projection
of the outputs of a run to $1$-receiving states must encode the run of
Turing machine $M$. As before, if the specification contains the
statement that ${\sf halt}$ must eventually be true, then the
specification is finite-state realizable if and only if $M$ halts.

Finally, we consider the isomorphic realizability problem instead of
the general distributed realizability problem. Since (for every $M$)
the given specification is such that any correct implementation for
one process can also be used for the other process, we can find a
solution for one problem if and only if we can find one for the
other.

As processes have no means of communication beyond
synchronization, the proof extends to rings of three or more
processes, where each additional process has the same specification as process $2$.
\qed


\medskip \noindent
Combining Theorems~\ref{thm:reduction} and \ref{thm:undecidable}, we obtain the
following result.

\begin{thm}
The parametric realizability problem is undecidable for
token rings and specifications of type b), c), or d).
\end{thm}

\proof
By Theorem~\ref{thm:reduction}, the isomorphic realizability problem
for a specification of type b) and (up to) three processes can be reduced to
a parameterized realizability problem of type b). Since the former
problem is undecidable by Theorem~\ref{thm:undecidable}, so is the
latter.  The proof for cases c) and d) is analogous.
\qed

Note that the proof of Theorem~\ref{thm:undecidable} does not work for
specifications of type a), since the specification relates outputs of one
process to outputs of the other. In fact, we can prove that the
parameterized realizability problem for type a) specifications is
decidable:

\begin{lem}
The parameterized realizability problem is decidable for token rings and
specifications of type a).
\end{lem}

\proof
This follows almost immediately from results of Clarke et
al.~\cite{Clarke04c} on token-passing networks.\footnote{See
  Section~\ref{sec:network} for details.} By their reduction, a
specification of the form $\forall i.\ \spec(i)$ holds for a process
implementation $\mT$ in a ring of arbitrary size if and only if $\spec(1)$ holds for
$\mT$ in a two-process system, where process $2$ has a fixed
implementation $\mT_2$ that does nothing but receive and (eventually) send the
token.

Furthermore, we can encode the behaviour of $\mT_2$ as additional 
assumptions in the specification, and let outputs of process $2$ be emulated by the environment.
Thus, we can synthesize a process implementation $\mT$ that satisfies a specification of
type a) in token rings of any size by defining assumptions on the
behavior of the token and synthesizing an implementation for one
process under these assumptions.\footnote{This synthesis approach for
  the parameterized synthesis of local specifications is mentioned as
  an optimization in Khalimov et al.~\cite{Khalimov13}.}
 That is, define assumption $A_\token$ as 
\[ A_\token \equiv \always(\neg \token_1 \impl \eventually \send_{2}) \land
\always(\token_1 \impl \neg \send_{2}), \]
and synthesize a process implementation $\mT$ satisfying
$ A_\token \impl \spec(1)$. \qed

\section{Bounded Isomorphic Synthesis}
\label{sec:bounded}

The reduction from Section~\ref{sec:synthesis} allows us to 
reduce parameterized synthesis to isomorphic synthesis
with a fixed number of processes. Still, the problem does not fall into
a class for which the distributed synthesis problem is decidable. 

For distributed architectures that do not fall into decidable classes,
Finkbeiner and Schewe have introduced \emph{bounded
  synthesis}~\cite{Finkbeiner12a}, a semi-decision procedure
that converts an undecidable 
distributed synthesis problem into a sequence of decidable synthesis problems,
by bounding the size of the implementation. In the following, we will show how
to adapt bounded synthesis for isomorphic synthesis in token rings, which by
Corollary~\ref{cor:reduction} amounts to parameterized synthesis in 
token rings.

\subsection{Bounded Synthesis}
The bounded synthesis procedure consists of three main steps:

\subsubsection*{Step 1: Automata translation.}
 Following an approach by Kupferman and Vardi~\cite{KupfermanV05}, the
  LTL specification $\spec$ (including fairness assumptions
like fair scheduling) is translated into a universal co-\Buchi automaton $\mU$ that
accepts an LTS $\mT$ if and only if $\mT$ satisfies $\spec$.

\subsubsection*{Step 2: SMT Encoding.}
Existence of an LTS that satisfies $\spec$ is encoded into a set of
SMT constraints over the theory of integers and free function symbols. States of
the LTS are represented by natural numbers in the bounded range $T={1, \ldots, k}$, 
state labels as free functions of
type $T \impl \bbB$, and the global transition function as a free function of
type $T \times \bbB^{|O_{env}|} \impl T$. Transition functions of
individual processes are defined indirectly by introducing projections $d_i:
T \impl T$, mapping global to local states. To ensure that local
transitions of process $i$ only depend on inputs in $I_i$, we add a constraint
\[ \bigwedge_{i \in P^-} ~ \bigwedge_{t,t' \in T} ~ \bigwedge_{I,I' \in \mathcal{P}(O_{env})}~ d_i(t) = d_i(t') \land I \cap I_i = I'
\cap I_i \impl d_i(\delta(t,I)) = d_i(\delta(t',I')).\]
To obtain an interpretation of these symbols that satisfies the
specification $\spec$, additional annotations of states are
introduced.  This includes labels $\lambda^\bbB_q: T \impl \bbB$
and free functions $\lambda^\#_q: T \impl \bbN$, defined
such that (i) $\lambda^\bbB_q(t)$ is true if and only if $(q,t) \in Q \times T$
is reachable in a run of $\mU$ on $\mT$\footnote{That is, there is a
  run of $\mU$ on $\mT$ such that at some point $q$ is among the
  states of $\mU$ when it reads (the label of) state $t$ of $\mT$.},
and (ii) valuations of the 
$\lambda^\#_q$ must be increasing along paths of $\mU$, and strictly
increasing for transitions that enter a rejecting state of
$\mU$. Together, this ensures that an LTS satisfying these constraints
cannot have runs which enter rejecting states infinitely often (and
thus would be rejected by $\mU$).


\subsubsection*{Step 3: Solving, Iteration for Increasing Bounds.}
The SMT constraints that result from step $2$ are in the theory of linear integer arithmetic with free function symbols. They are decidable because the number of processes, the size of process implementations and the number of inputs to each process are bounded. If any of these were unbounded, we would have to use unbounded quantification instead of the finite conjunction in step $2$, making the satisfiability problem undecidable.

Thus, for a given bound $k$ on the size of $T$, we can decide satisfiability of the constraints. If the constraints are unsatisfiable for a given bound $k$, we increase $k$, add the necessary formulas to the encoding, and try again. If they are satisfiable, we obtain a model,
giving us an implementation for the system processes such that $\spec$ is
satisfied.

\begin{thm}[\cite{Finkbeiner12a}]
If a given LTL specification $\spec$ is realizable in a given architecture $A$, then the bounded synthesis procedure will eventually terminate and return implementations of the system processes that satisfy $\spec$ in $A$.
\end{thm}

\begin{exa}
As a very simple example with just one process $P$, consider the
specification $\always(r \rightarrow \eventually g)$, where $r$ is an
input and $g$ an output variable of
$P$. Figure~\ref{fig:bounded-example1} depicts the resulting universal
co-\Buchi automaton, and Figure~\ref{fig:bounded-example2} the resulting
set of SMT constraints.\footnote{Note that in the automaton, for the
  sake of brevity we use the notation $r\overline{g}$ instead of $r
  \land \neg g$.}

The constraints encode, from top to bottom, annotations corresponding
to states and transitions of $\mU$. In particular, we have annotations
for i) the initial state of $\mT$ (and $\mU$), ii) states reachable by
any transition from a state $t$ with $\lambda_0^\bbB(t)$, iii) states
reachable by a transition with $r \wedge \neg g$ from a state $t$ with
$\lambda_0^\bbB(t)$, and iv) states reachable by a transition with
$\neg g$ from a state $t$ with $\lambda_1^\bbB(t)$. 
\end{exa}

\begin{figure}
\begin{pspicture}(0,0.5)(4,2)
\psset{arrows=->,nodesep=5pt}
\put(0,1){\rnode{emt}{}}
\cnodeput(2,1){0}{$0$}
\cnodeput(4,1){1}{$1$}
\put(4,1){\Cnode[radius=0.35cm]{a}}

\ncline{0}{1}\taput{$r\overline{g}$}
\nccircle[angle=0,nodesep=2pt]{0}{0.25cm}\taput{$*$}
\ncline{emt}{0}
\nccircle[angle=0,nodesep=2pt]{1}{0.25cm}\taput{$\overline{g}$}

\end{pspicture}
\caption{Universal co-\Buchi automaton $\mU$ for $\always(r \rightarrow \eventually g)$}
\label{fig:bounded-example1}
\end{figure}

\begin{figure}
$\begin{array}
{ll}
& \lambda^\bbB_0 (0) \land \lambda^\#_0(0)=0\\
\bigwedge_{t \in T}~\bigwedge_{I \in \mathcal{P}(O_{env})}~& \lambda^\bbB_0(t) \impl \lambda^\bbB_0(\delta(t,I)) \land
\lambda^\#_0(\delta(t,I)) \geq \lambda^\#_0(t)\\
\bigwedge_{t \in T} \bigwedge_{I \in \mathcal{P}(O_{env})}~& \lambda^\bbB_0(t) \land r
\in I \land \neg g(t) \impl \lambda^\bbB_1(\delta(t,I)) \land \lambda^\#_1(\delta(t,I)) > \lambda^\#_0(t)\\
\bigwedge_{t \in T} \bigwedge_{I \in \mathcal{P}(O_{env})} &
\lambda^\bbB_1(t) \land \neg g(t) \impl \lambda^\bbB_1(\delta(t,I)) \land
\lambda^\#_1(\delta(t,I)) > \lambda^\#_1(t)
\end{array}$
\caption{SMT constraints for $\always(r \rightarrow \eventually g)$}
\label{fig:bounded-example2}
\end{figure}

\subsection{Adaption to Token Rings}
\label{sec:adapted}

We adapt the bounded synthesis approach for synthesis in token rings, and
introduce some optimizations we found vital for a good performance of the
synthesis method.

\subsubsection*{Additional Constraints and Optimizations.}
We use some of the general modifications and optimizations mentioned
in Finkbeiner and Schewe~\cite{Finkbeiner12a}:
\begin{itemize}
\item We modify the constraints to ensure that the resulting system
  implementation is asynchronous. In general (see
  Section~\ref{sec:prelim-distributed}), we could directly add a scheduling
  variable $s_i$ for each process $i$ and a constraint 
	\[ \bigwedge_{i \in P^-}~ \bigwedge_{I \in \mathcal{P}(O_{env})}~s_i \not \in I \impl d_i(\delta(t,I)) =
  d_i(t). \] 
For the synthesis of token rings we use a modified version, explained below.
\item We use symmetry constraints to encode that all processes should be
  isomorphic. Particularly, we use the same
  function symbols for state labels of all system processes, and special
  constraints for the local transition functions, also explained below.
\item We use the semantic variant where environment inputs are not stored in system
states, but are directly used in the transition term that computes the following
state. This results in an
implementation that is a factor of $\card{O_{env}}$ smaller.\footnote{The
  different semantics (compared to the input-preserving LTSs used
  in~\cite{Finkbeiner12a}) is already reflected in our definition of LTSs and
  satisfaction of LTL formulas.}
\item Finally, we use real numbers instead of integers as the codomain of
  functions $\lambda^\#_q$, as real arithmetic can be solved more efficiently.
\end{itemize}

\subsubsection*{Encoding Token Rings.}
For the synthesis of token rings, we use the following modifications to
the SMT encoding:
\begin{itemize}
\item We want to obtain an asynchronous system in which the environment is
  always scheduled, along with exactly one system process. Thus, we do
  not need $|P|$ scheduling variables, but can encode the index of the scheduled
  process into a binary representation with $log_2(|P^-|)$ inputs.
\item We encode the special features of token rings:
  \begin{enumerate}[label=\roman*)]
\item exactly one process should have the token at any time,
\item only a process that has the token can send it,
\item if process $i$ wants to send the token, and process $i+1$ is scheduled, 
	then in the next state process $i+1$ has the token and process
  $i$ does not,
\item if process $i$ has the token and does not send it (or process $i+1$ 
	is not scheduled), it also has the token in the next state,
        and
\item if process $i$ 
	does not have the token and does not receive it from process $i-1$, then it will also 
	not have the token in the next step.
\end{enumerate}
Properties ii) -- v) are encoded in
  the following constraints, 
  where 
\begin{itemize}
\item $\token_i(d_i(t))$ is
  true in state $t$ if and only if process $i$ has the token, 
\item $\send(d_i(t))$ is true if and only if
  $i$ is ready to send the token, and 
\item $\sched_i(I)$ is true if and only if the scheduling
  variables in $I$ are such that process $i$ is scheduled.
\end{itemize}
 
	\begin{align*}
  \bigwedge_{i \in P^-}~\bigwedge_{t \in T}~ \hspace{48pt}&
	\neg \token(d_i(t)) \impl \neg \send(d_i(t))\\[5pt]	
  \bigwedge_{i \in P^-}~\bigwedge_{t \in T}~\bigwedge_{I \in \mathcal{P}(O_{env})}~&
	\send(d_i(t)) \land \sched_{i+1}(I) \impl \neg \token(d_i(\delta(t,I)))\\[5pt]	
  \bigwedge_{i \in P^-}~\bigwedge_{t \in T}~\bigwedge_{I \in \mathcal{P}(O_{env})}~&
	\send(d_{i-1}(t)) \land \sched_i(I) \impl \token(d_i(\delta(t,I)))\\[5pt]	
  \bigwedge_{i \in P^-}~\bigwedge_{t \in T}~\bigwedge_{I \in \mathcal{P}(O_{env})}~&
	\token(d_i(t)) \impl (\send(d_i(t)) \land \sched_{i+1}(I)) \lor \token(d_i(\delta(t,I)))\\[5pt]	
	\bigwedge_{i \in P^-}~\bigwedge_{t \in T}~\bigwedge_{I \in \mathcal{P}(O_{env})}~&
	\neg \token(d_i(t)) \land \neg (\send(d_{i-1}(t)) \land
        \sched_i(I))\impl \neg \token(d_i(\delta(t,I))). \hspace{-34pt}
  \end{align*}
  We do not encode property
  i) directly, because it is implied by the remaining constraints whenever we
  start in a state where only one process has the token. Without loss
  of generality, we can assume that process $1$ initially has the
  token, expressed as
\[ \token(d_1(0)) \land \bigwedge_{i \in P^-\setminus\{1\}} \neq
\token(d_i(0)). \]
\item Token passing is an exception to the rule that only the scheduled
  process changes its state: if process $i$ is scheduled in state $t$, and 
	$\send(d_{i-1}(t))$ holds, then in the following transition
  both processes $i-1$ and $i$ will change their state. The constraint that
  ensures that only scheduled processes may change their state is modified into

\[\bigwedge_{i \in P^-}~\bigwedge_{t \in T}~\bigwedge_{I \in \mathcal{P}(O_{env})}~ 
	\neg \sched_i(I) \land \neg (\send(d_i(t)) \land \sched_{i+1}(I))
	\impl d_i(\delta(t,I)) = d_i(t).\hspace{-36pt} \]
	
	\item Finally, we need to restrict local transitions in order to obtain
  isomorphic processes. The general rule is that local transitions of process
  $i$ should be determined by the local state and inputs in $I_i$. With our
  definition, token passing is an exception to this rule. The resulting
  constraints for local transitions are:
\begin{align*}
  \bigwedge_{i \in P^-\setminus\{1\}}~\bigwedge_{t,t' \in T}~\bigwedge_{I,I' \in \mathcal{P}(O_{env})}~&
	\left( \begin{array}{l} 
					d_1(t) = d_i(t') \land \sched_1(I) \land \sched_i(I') \land I \cap I_1 = I' \cap I_i\\[5pt]
					\impl d_1(\delta(t,I)) = d_i(\delta(t',I')) 
					\end{array} \right) \hspace{-36pt}  \\[5pt]
	\bigwedge_{i \in P^-\setminus\{1\}}~\bigwedge_{t,t' \in T}~\bigwedge_{I,I' \in \mathcal{P}(O_{env})}~&
  \left( \begin{array}{l}
					d_1(t) = d_i(t') \land \send(d_1(t)) \land \send(d_i(t')) \\[5pt]
					\land~ \sched_2(I) \land \sched_{i+1}(I') \land I \cap I_1 = I' \cap I_i \\[5pt]
					\impl d_1(\delta(t,I)) = d_i(\delta(t',I'))
				 \end{array} \right). \hspace{-36pt} \\	
  \end{align*}
	
\end{itemize}

\subsubsection*{Fairness of Scheduling and Token Passing.}
A precondition of Theorem~\ref{thm:reduction} is that the
  implementation needs to ensure fair token-passing. Thus, we always add
\[ \forall i.~{\sf fair\_scheduling} \impl \always (\token_i \impl
\eventually \send_i) \]
 to $\spec$, where ${\sf fair\_scheduling}$ stands for $\forall j.~\always \eventually
\sched_j$. Note that with this condition, the formula does not fall
into any of the cases from Theorem~\ref{thm:reduction}. However, by adding this formula we 
only make explicit the assumption of fair token passing, which obviously necessitates fair 
scheduling. Thus, this formula does not need to be taken into account when choosing which 
of the cases of the theorem needs to be applied.

Similarly, the ${\sf fair\_scheduling}$ assumption needs to be added to
  any liveness conditions of the specification, as without fair scheduling in
  general liveness conditions cannot be guaranteed. As before, this does not
  need to be taken into account considering Theorem~\ref{thm:reduction}.


\subsubsection*{Correctness and Completeness of Bounded Synthesis for Token Rings.}

Based on correctness of the original bounded synthesis approach (and correct
modeling of the features of token rings), we obtain

\begin{cor}
\label{cor:synthesis:isomorphic}
If a given specification $\spec$ is realizable in a token ring of a given
size $n$, then the bounded synthesis procedure, adapted to token rings, will
eventually find this implementation.
\end{cor}
Since we have shown in Theorem~\ref{thm:undecidable} that the isomorphic realizability 
problem is undecidable in token rings, there is no algorithm that can
also detect unrealizability in all cases. In fact, the given procedure
will not terminate if the specification is unrealizable.

Finally, based on the correctness of our adaption of bounded synthesis,
and Corollary~\ref{cor:reduction}, we obtain 

\begin{thm}
\label{cor:synthesis:reduction}
If a given specification $\pspec = \forall
\overline{t}.\ \spec(\overline{t})$ falls into class a) (b,c,d) of
Theorem~\ref{thm:reduction} and the adapted bounded synthesis
algorithm finds a process implementation $\mT$ such that, for a
parameteric token ring $\Pi_R$ and $m=2$ ($3$,$4$,$5$),

\[ \textrm{for } 1 \leq k \leq m:\ \Pi_R(k), \mT \models \bigwedge_{\substack{1 \leq i,j \leq
    k\\ i \neq j}} \spec(\overline{t}),\]
then $\mT$ satisfies $\pspec$ in token rings of arbitrary size.
\end{thm}

\section{Network Decomposition for General Token-passing Systems}
\label{sec:network}

Clarke, Talupur, Touilli, and Veith \cite{Clarke04c} have extended the results
of Emerson and Namjoshi to 
arbitrary token-passing networks. They reduce the parameterized verification
problem to a finite set of model checking problems, where the number of problems
and the size of systems to be checked depends on the architecture of the
parameterized system and on the property to be proved. In the following, we 
recapitulate their results and show how they can be applied to allow for 
synthesis of processes in general token-passing networks.

\newpage
\subsection{Definitions}

\subsubsection*{Network Graph} A \emph{network graph} is a finite directed graph 
$G=(S,C)$ without self-loops, where $S$ is the set of \emph{processes}, and $C$ is the 
set of \emph{connections}. A \emph{path} in $G$ is a sequence of processes $s_1 s_2 \ldots s_n$ such that for $1 \leq i < n$, $(s_i,s_{i+1}) \in C$. A path is \emph{$R$-free} for a set $R \subseteq S$, if $s_i \notin R$ for all $s_i$ with $1 < i < n$.

\subsubsection*{Token-passing Network} We consider \emph{token-passing
  networks} based on network graphs. Let $I_i$ be isomorphic sets of
indexed input variables for all processes $i \in S$, with $\send_j \in
I_i$ for at least one $j\neq i \in S$. Similarly, let $O_i \supseteq
\{\token_i, \send_i\}$ be isomorphic sets of output variables for all
processes. Let furthermore $env$ be the environment process with
outputs $O_{env}$, such that $(\bigdotcup_{i \in S} O_i) \cap O_{env}
= \emptyset$, and let $P=S \cup \{env\}$. Together with this
\emph{interface} for the processes, a network graph $G$ corresponds to
the architecture 
\[A_G = (P, env, \bigcup_{i \in P} O_i, \{I_i \mid i \in P^-\}, \{O_i \mid i \in P\}).\]
Note that in contrast to token rings, we may have several connections that allow sending or 
receiving the token for each process, i.e., we may have $\send_i \in I_j$ for more than one $j \in S$, and $\send_k \in I_i$ for more than one $k \in S$. The decision which of these connections is used is left to the scheduler, i.e., the environment process $env$: if $\send_i$ is active, then the next process $j$ with $\send_i \in I_j$ that is scheduled will receive the token. Similar to the case of token rings, we consider only networks and schedulers that ensure fair token passing, i.e. in every execution of the system, every process will receive the token infinitely often.

\begin{figure}
\scalebox{0.16}{\includegraphics{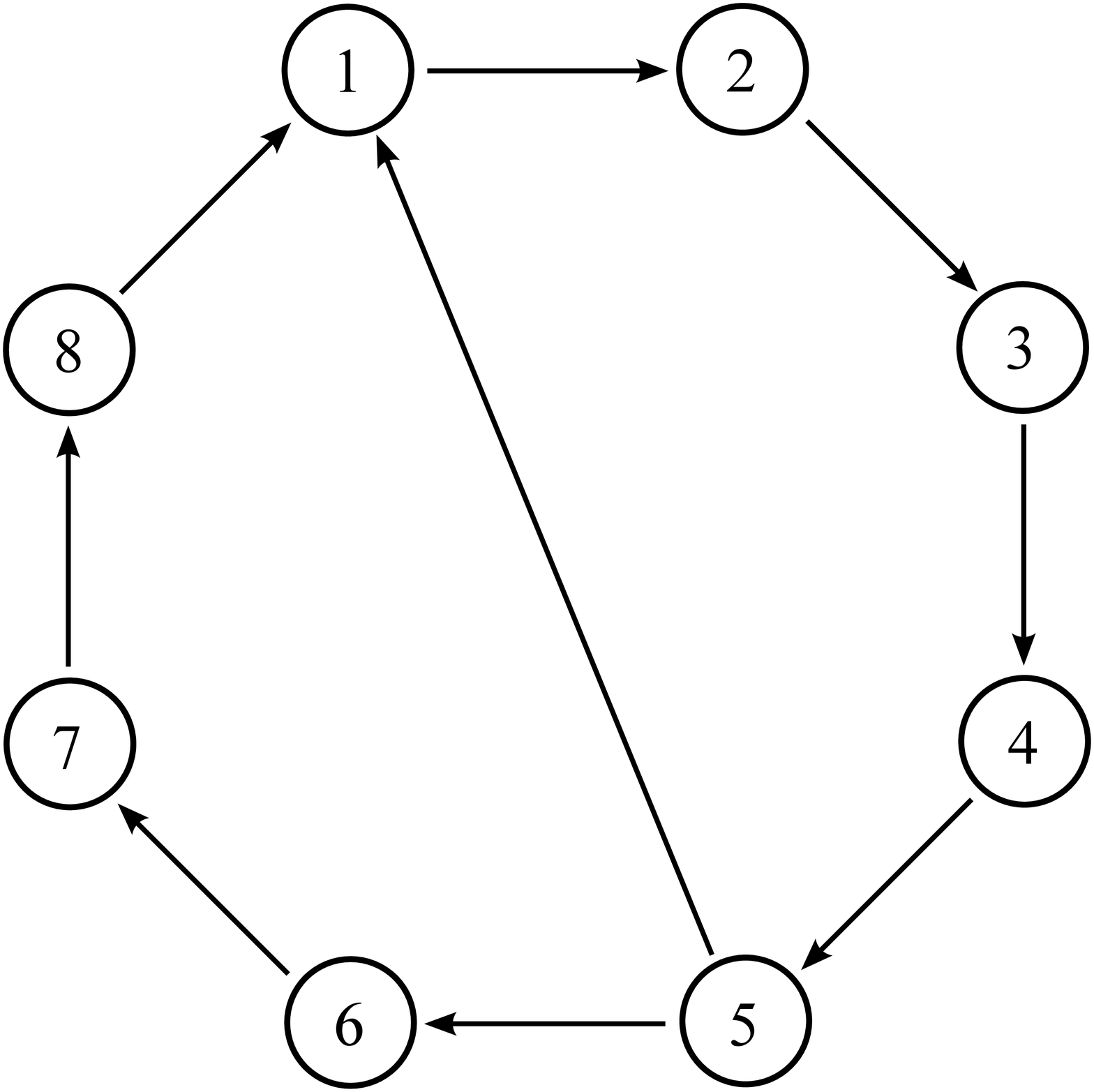}}
\caption{Network Graph for a Prioritized Token Ring}
\label{fig:PrioArbiter}
\end{figure}

\begin{exa}[Token-Passing Network]
\label{ex:PrioArbiter}
Figure~\ref{fig:PrioArbiter} shows the network graph $G^{\sf prio}$ that resembles a token ring, except that there is an additional ``shortcut'' connection between processes $5$ and $1$. $G^{\sf prio}$ can be seen as a token ring with additional prioritization: whenever the token is passed by process $5$, the environment can decide whether the low-priority processes on the left-hand side will receive the token in this round (by scheduling process $6$) or not (by scheduling process $1$). The fairness assumptions ensure that every process will receive the token infinitely often.
\end{exa} 

\subsubsection*{k-Indexed Formula} A \emph{$k$-indexed formula} is a formula with 
arbitrary quantification in prenex form that refers to at most $k$ different processes, 
i.e., there are at most $k$ different constant indices and index variables.

\subsubsection*{Connectivity, Connection Topology} Consider a network graph $G = (S,C)$ and a subset $R = \{p_1,\ldots,p_k\} \subseteq S$ of processes. We define the following connectivity properties for index variables $x,y$:

\medskip
\begin{tabular}{ll}
$G \models \Phi_\circlearrowleft(x,R)$ & ``There is an $R$-free path from $x$ to itself''\\
$G \models \Phi_\leadsto(x,y,R)$ & ``There is a path from $x$ to $y$ via a third process not in $R$''\\
$G \models \Phi_\rightarrow(x,y)$ & ``There is a direct connection $(x,y) \in C$''
\end{tabular}

\medskip
\noindent By instantiating variables $x$ and $y$ with elements of $R$ in all possible combinations, we obtain a finite set of different conditions, describing all possible connectivities between processes in $R$. These connectivities represent the \emph{connection topology} of $G$ with respect to $R$, denoted $G_R$. The connection topology $G_R$ can be depicted in a network graph $G_R=(S_R,C_R)$ with at most $2k$ nodes, where $S_R$ contains \emph{sites} $site_1, \ldots, site_k$ corresponding to elements of $R$, and a number of ``hub'' nodes, each representing one or several nodes from $S \setminus R$. The minimal network graph with these properties can be used as a representative of the connection topology.

For a given connection topology $CT$ and process interface, we will
denote by $A_{CT}$ the architecture based on $CT$, where sites are
represented by processes as usual, and hubs are replaced by processes
with a fixed  implementation that always eventually passes on the
token.\footnote{Actually, an implementation with three states is
  sufficient: one where the process waits for the token, one state it
  enters when receiving the token, and one where it sends the
  token. The latter is always entered when it is scheduled for the
  first time after receiving the token. Since processes cannot observe
  how many steps the other processes take, this preserves full
  generality.} 

\begin{figure}
\scalebox{0.16}{\includegraphics{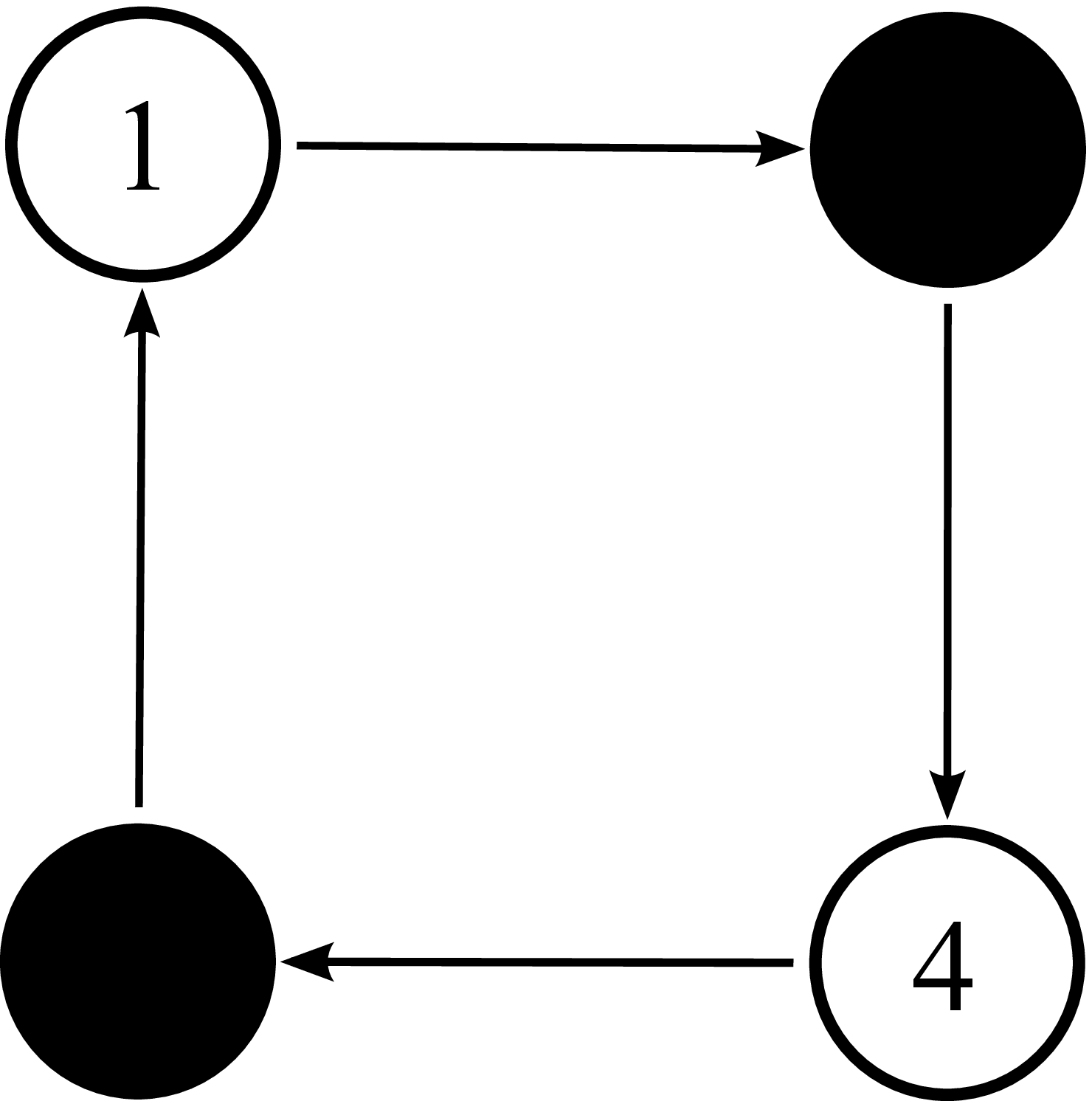}}
\caption{Connection Topology $G^{\sf prio}_{\{1,4\}}$}
\label{fig:ConTopo}
\end{figure}

\begin{exa}
Figure~\ref{fig:ConTopo} shows the connection topology $G^{\sf prio}_R$ of the network graph from Example~\ref{ex:PrioArbiter} with respect to $R = \{1,4\}$. Hub nodes are depicted as filled black circles.
\end{exa}

\subsubsection*{$k$-Topology} Given a network graph $G=(S,C)$, the $k$-topology of $G$ is
\[ CT_k(G) = \left\{ G_R \mid R \subseteq S, \card{R} = k \right\}. \]

\begin{figure}
\begin{tabular}{ccc}
\scalebox{0.16}{\includegraphics{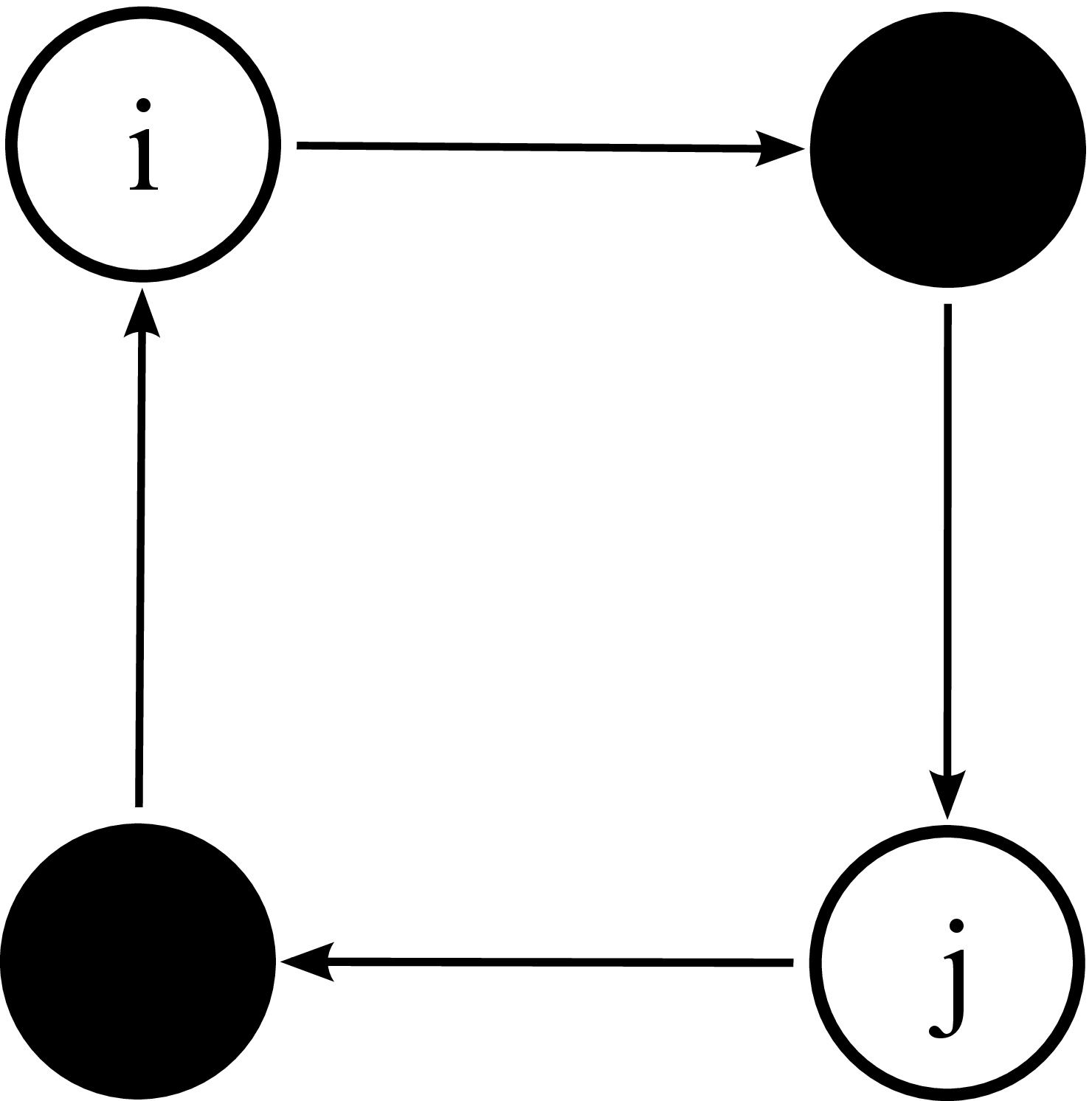}} & \scalebox{0.16}{\includegraphics{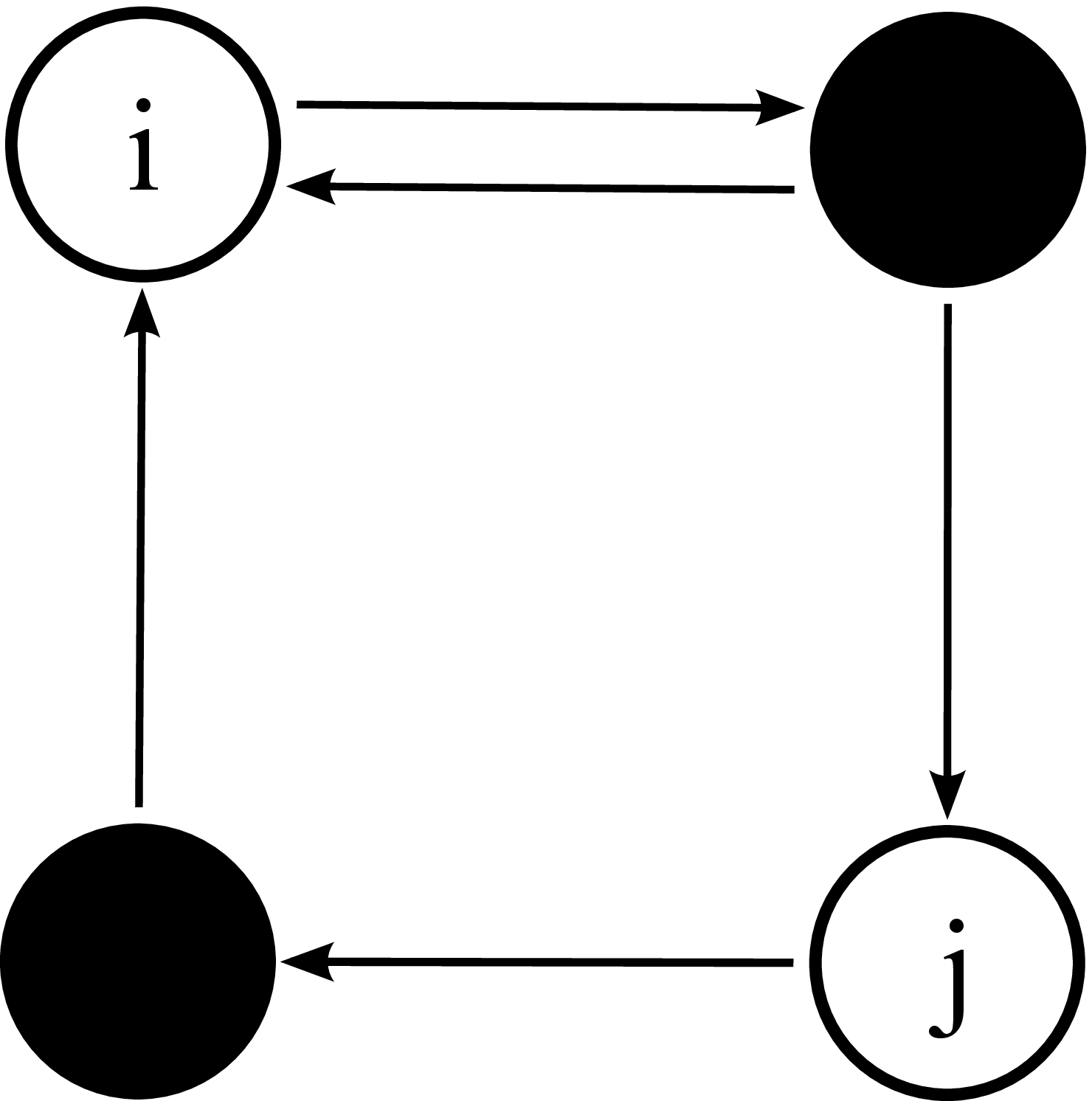}} & \scalebox{0.16}{\includegraphics{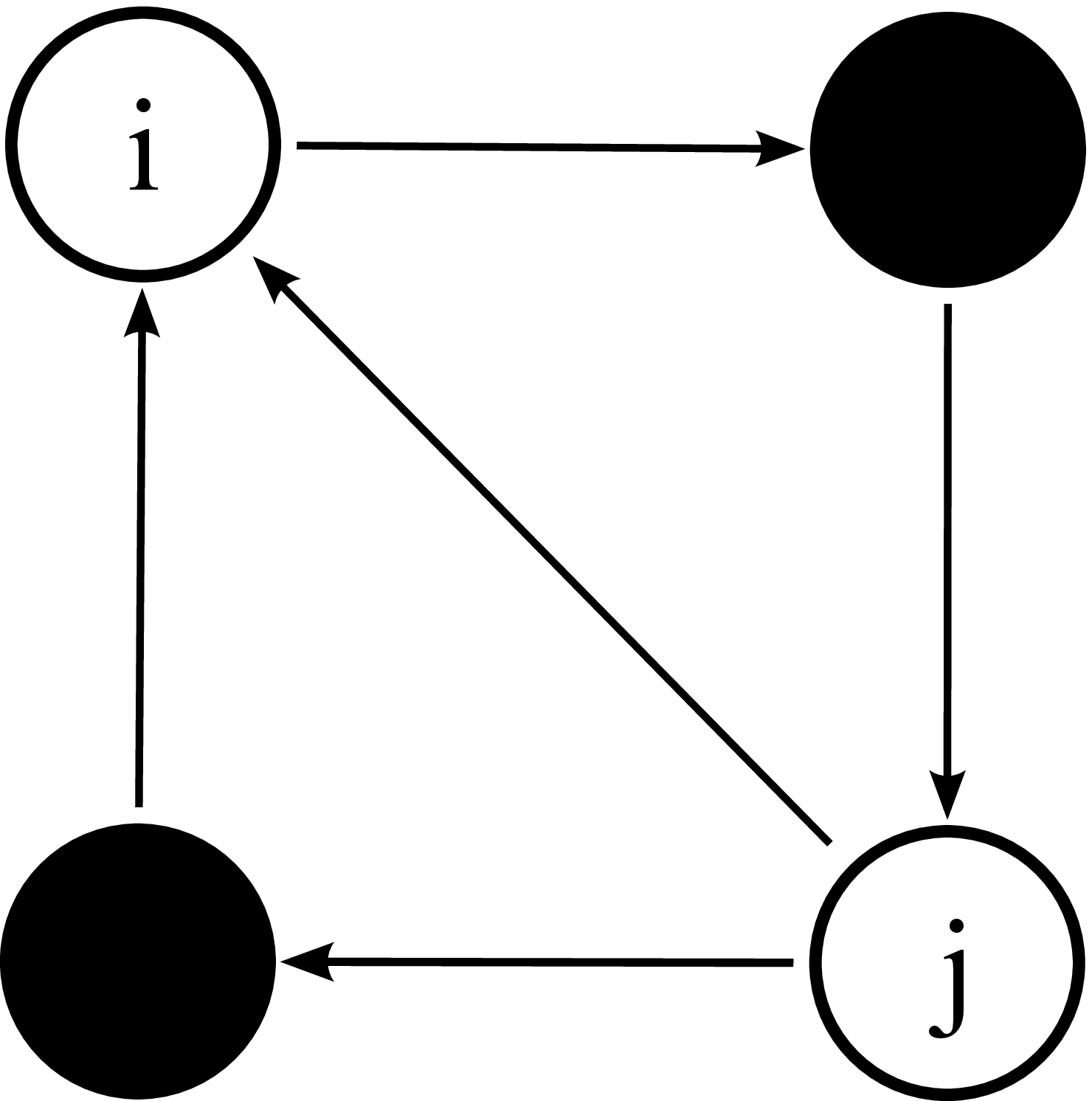}}\\
a) & b) & c)\\[20pt]
\scalebox{0.16}{\includegraphics{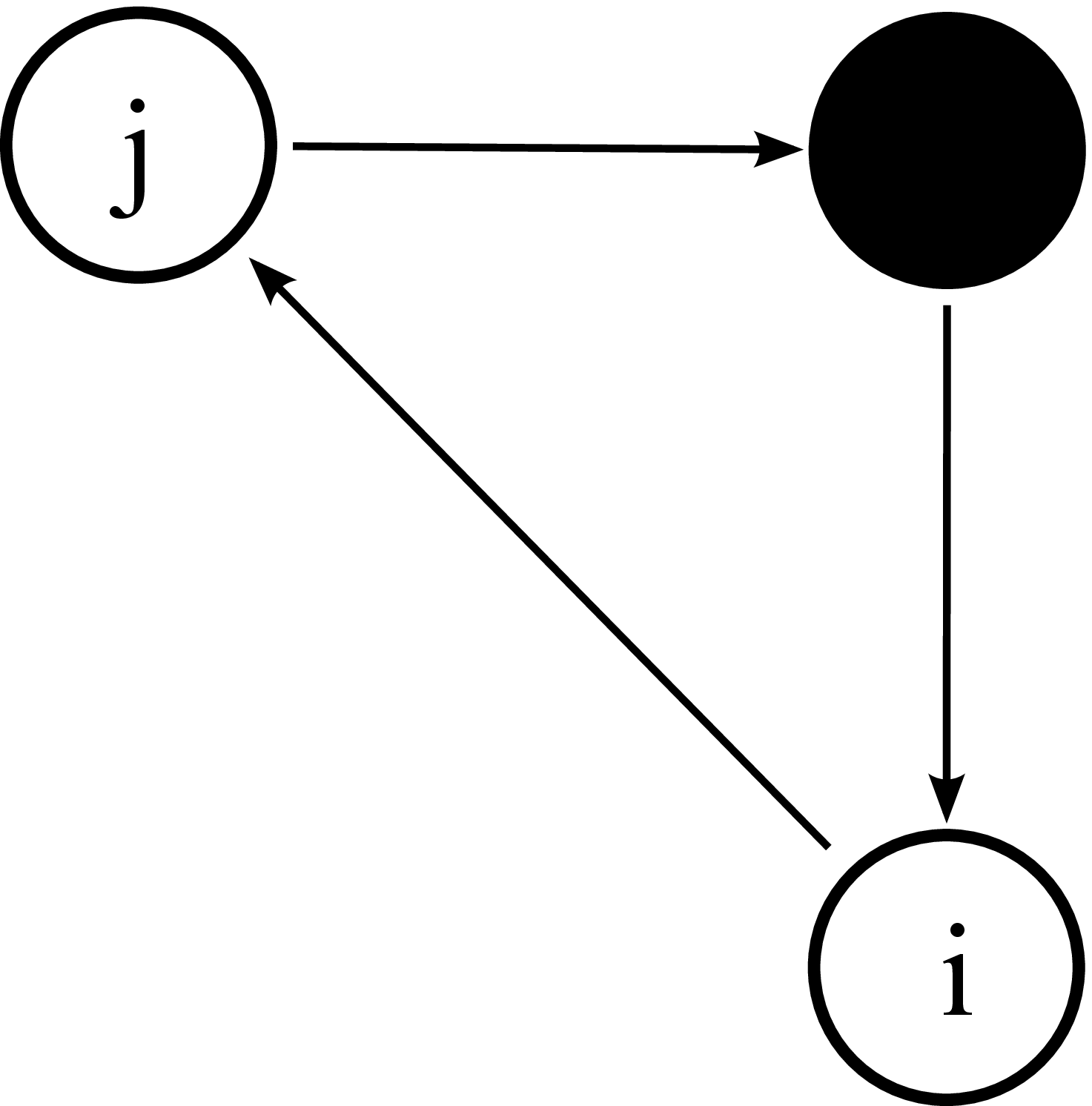}} & \scalebox{0.16}{\includegraphics{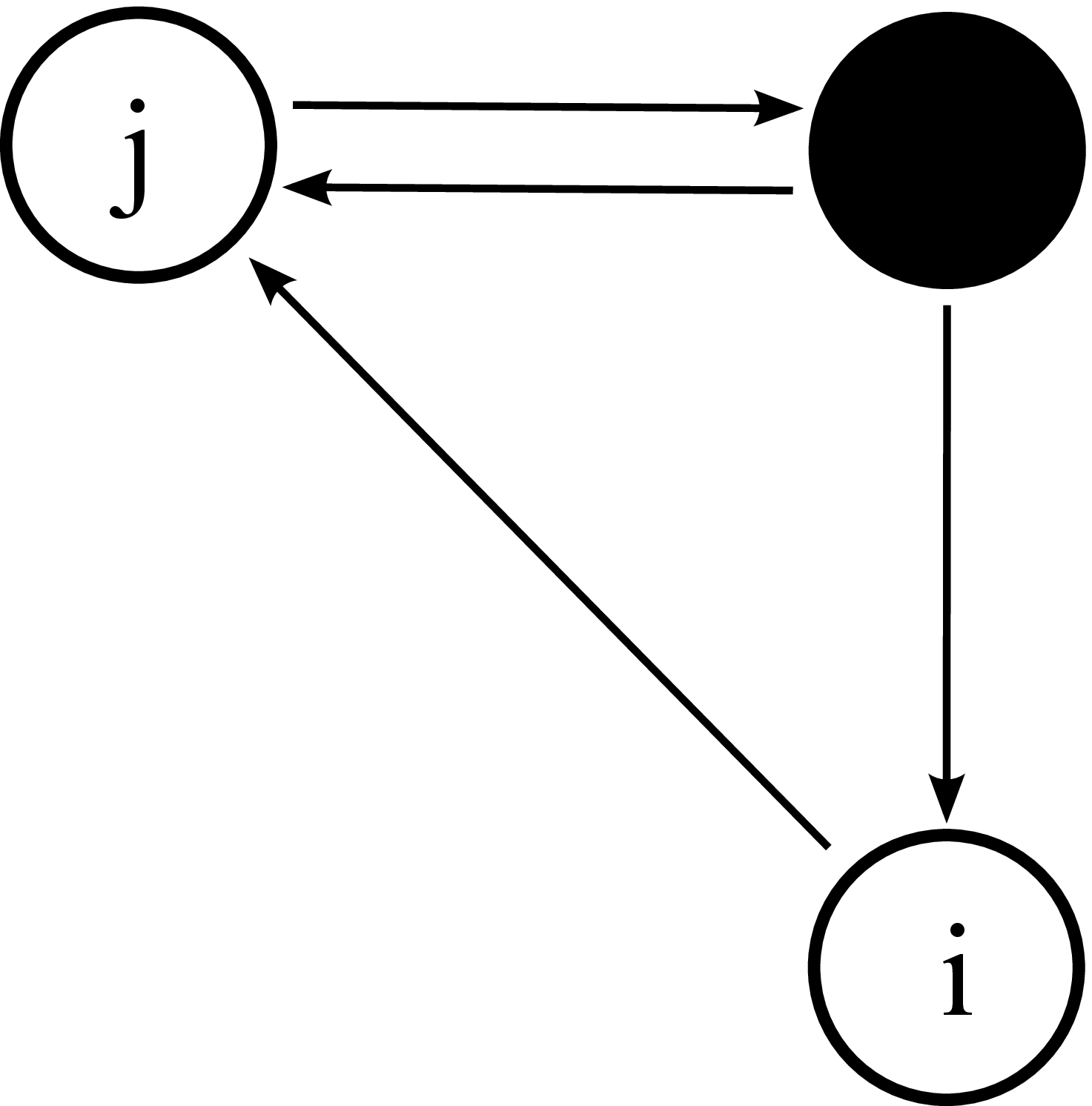}} & \scalebox{0.16}{\includegraphics{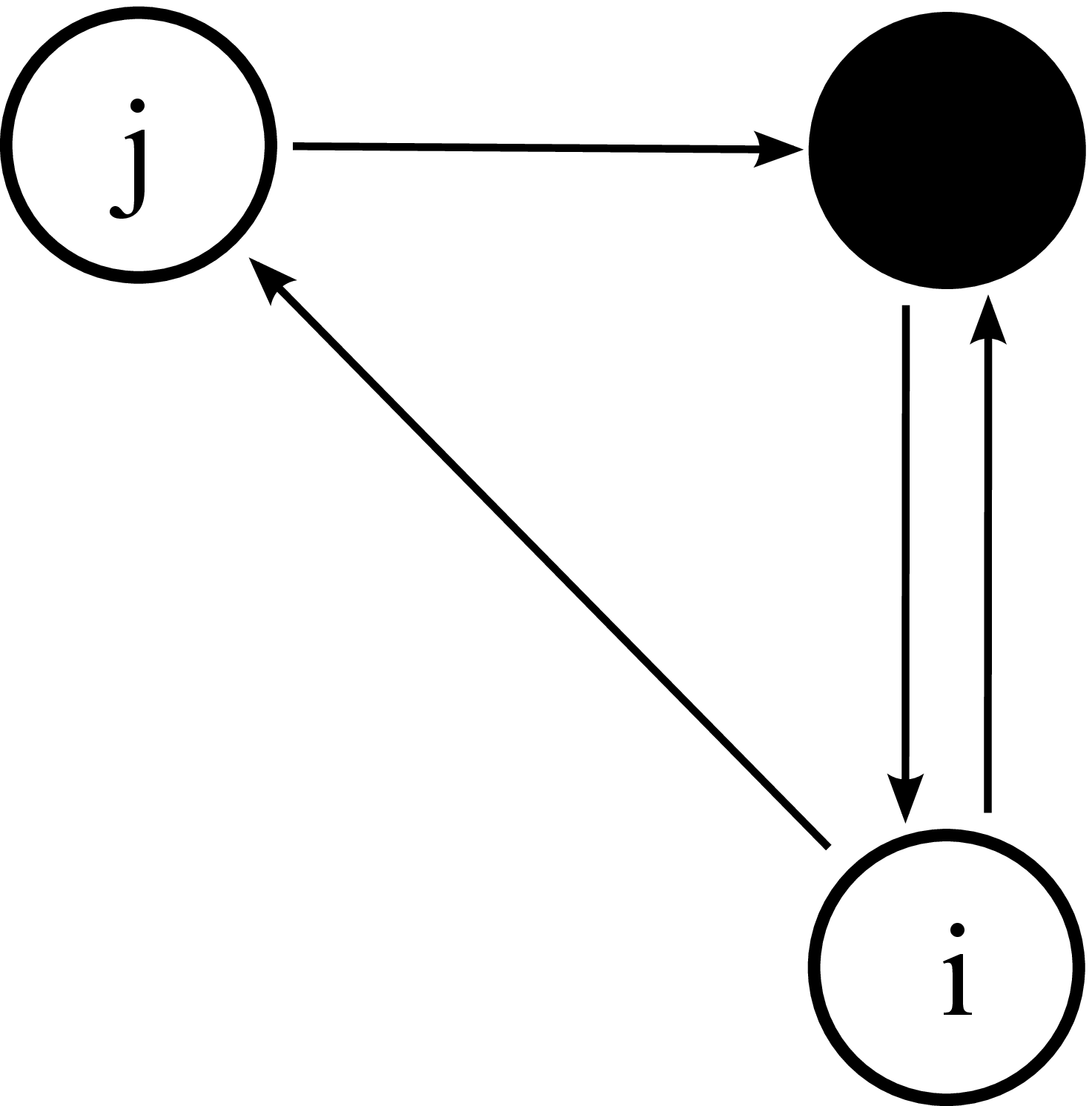}}\\
d) & e) & f)
\end{tabular}
\caption{2-Topology of $G^{\sf prio}$}
\label{fig:2Topo}
\end{figure}

\begin{exa}
\label{ex:2Topo}
Figure~\ref{fig:2Topo} shows the $2$-topology of the network graph
$G^{\sf prio}$ from Example~\ref{ex:PrioArbiter}. Note that several
subsets $R \subseteq S$ have the same topology. E.g., the topology in
a) is $G^{\sf prio}_{\{i,j\}}$ for any two processes $i,j \in S$ which
are both high- or low-priority, and are not neighbors. Also, for all
topologies there are symmetric variants, where $i$ and $j$ switch
positions. Except for a), the symmetric variants are different from
the original topology. We do not depict these variants. 
\end{exa}

\subsection{Verification and Synthesis by Network Decomposition}

The main result of Clarke et al.~\cite{Clarke04c} allows for model checking $k$-indexed properties in a given network graph $G$ by reduction to model checking in the $k$-topology of $G$:

\begin{thm}[\cite{Clarke04c}]
\label{thm:ND}
Let $\mT$ be a process implementation, $A_G$ an architecture based on
network graph $G$ and the interface of $\mT$, and $\exists
\overline{i}.~\spec(\overline{i})$ a $k$-indexed formula. Then
\[ A_G, \mT \models \exists~ \overline{i}.~ \spec(\overline{i})  ~~\Longleftrightarrow~~ \exists~ CT \in CT_k(G).~ A_{CT}, \mT \models \spec(site_1,\ldots,site_k)\]
\end{thm}

This result can be extended to a model checking approach for arbitrary combinations of quantifiers by rewriting universal (existential) quantifiers into explicit conjunctions (disjunctions) over all connection topologies in $CT_k(G)$, and checking $A_{CT}, \mT \models \spec(site_1,\ldots,site_k)$ for all $CT \in CT_k(G)$ to evaluate the resulting formula. That is, for a $k$-indexed formula $\overline{Q} \overline{i}.~ \spec(\overline{i})$ with arbitrary quantifier prefix $\overline{Q} \overline{i}$, we rewrite the model checking problem $A_G, (\mT,\ldots,\mT) \models \overline{Q} \overline{i}.~ \spec(\overline{i})$ until saturation, according to the following rules:

\begin{align*}
A_{G_R}, \mT \models \forall i~\overline{Q}' \overline{i}'.~\spec(\ldots,i,\ldots) & ~\mapsto~ \bigwedge_{i \in S\setminus R} A_{G_{R\cup \{i\}}}, \mT \models \overline{Q}' \overline{i}'.~\spec(\ldots,i,\ldots)\\
A_{G_R}, \mT \models \exists i~\overline{Q}' \overline{i}'.~\spec(\ldots,i,\ldots) & ~\mapsto~ \bigvee_{i \in S\setminus R} A_{G_{R\cup \{i\}}}, \mT \models \overline{Q}' \overline{i}'.~\spec(\ldots,i,\ldots),
\end{align*}
where $A_G = A_{G_\emptyset}$ and $\overline{Q}' \overline{i}'$ is a (possibly empty) quantifier prefix over the remaining index variables. Upon saturation, we obtain a Boolean combination of model checking problems $A_{CT}, \mT \models \spec(\overline{i})$, for all possible topologies $CT \in CT_k(G)$, where $\spec(\overline{i})$ is quantifier-free and the elements of $\overline{i}$ are instantiated to concrete elements of $S$. Since several subsets $R \subset S$ have the same topology $G_R \in CT_k(G)$, we do not need to solve $\card{S}^k$ model checking problems, but only as many as there are different topologies in $CT_k(G)$.

\subsubsection*{Reductions in Token-Passing Networks}
With some restrictions, the result above provides a reduction from the
parameterized model checking problem to a set of finite-state model
checking problems for a given class $\mG$ of network graphs with
$k$-topology $CT_k$:
\begin{itemize}
\item if the quantifier prefix is purely universal, then we can check
  validity of $\forall \overline{i}.~\spec(\overline{i})$ in all
  graphs of the class by checking whether
  $\spec(\overline{i})$ holds for all topologies in $CT_k$. 
\item similarly, we can check $\exists \overline{i}.~\spec(\overline{i})$ 
by checking whether $\spec(\overline{i})$ holds for at least one topology in $CT_k$.
\item if we have quantifier alternations, the problem is not so simple. In general, to define a
  reduction for a class of network graphs and a formula $\overline{Q}
  \overline{i}.~\spec(\overline{i})$, we 
  additionally need a Boolean function $B$ over variables whose truth
  values are defined by model checking elements of the
  $k$-topology. 
	That is, to check whether $\overline{Q}
  \overline{i}.~\spec(\overline{i})$ holds in all network graphs of
  the class, we let $CT_k = \{ CT^1, \ldots, CT^m \}$, define valuations of Boolean
  variables $g_j$ for $1 \leq j \leq m$ by $g_j \Leftrightarrow CT^j,
  \mT \models \spec(\overline{i})$, and evaluate $B(g_1,\ldots,g_m)$. 
Clarke et al.~\cite{Clarke04c} prove that for every network topology
and $k$-indexed quantifier prefix $\overline{Q} \overline{i}$ there is
a reduction $(CT_k, B(g_1,\ldots,g_m))$, but do not show
how to find $B$.
\end{itemize}

\begin{figure}
\begin{tabular}{cc}
\scalebox{0.16}{\includegraphics{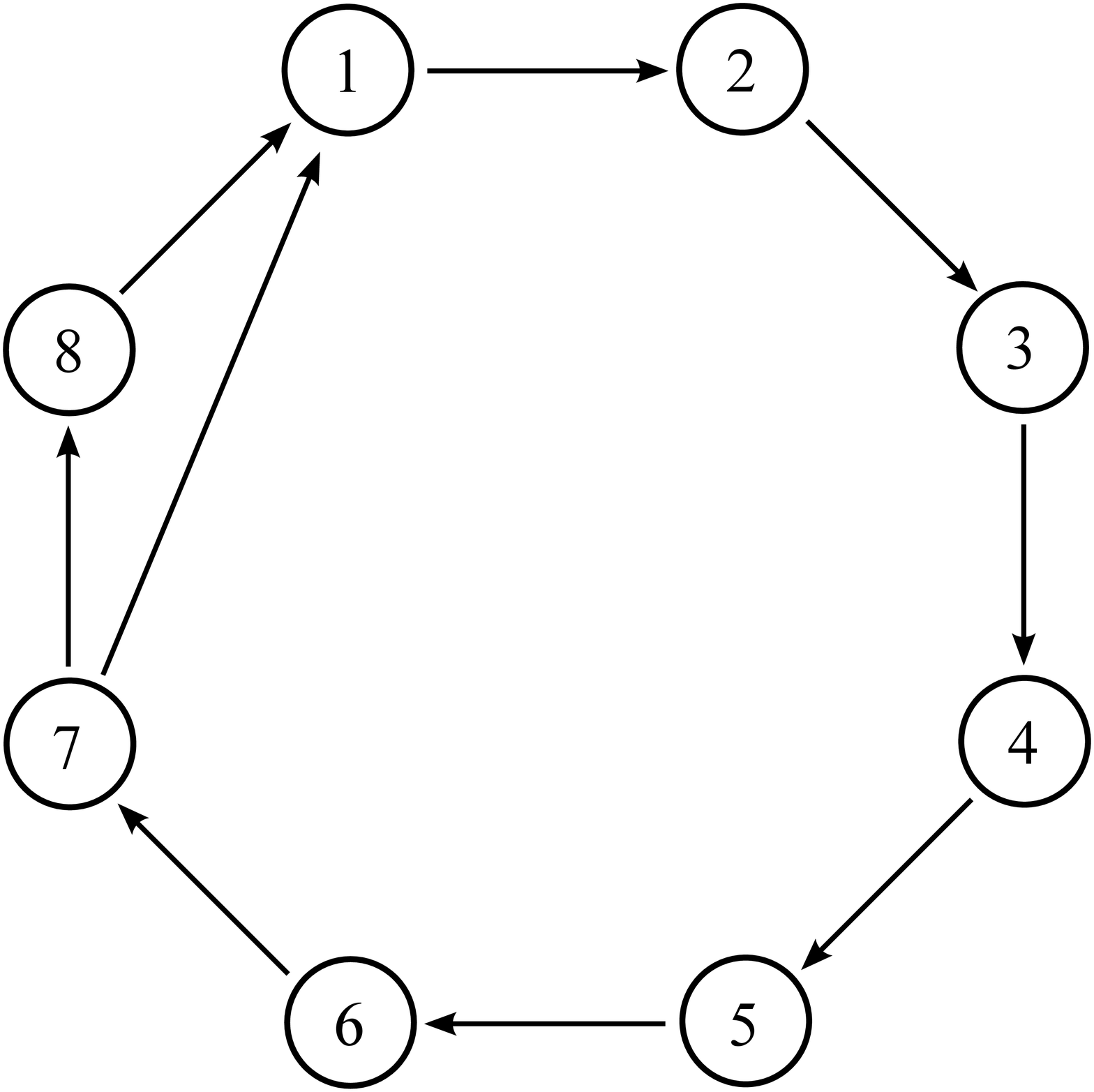}} & \scalebox{0.16}{\includegraphics{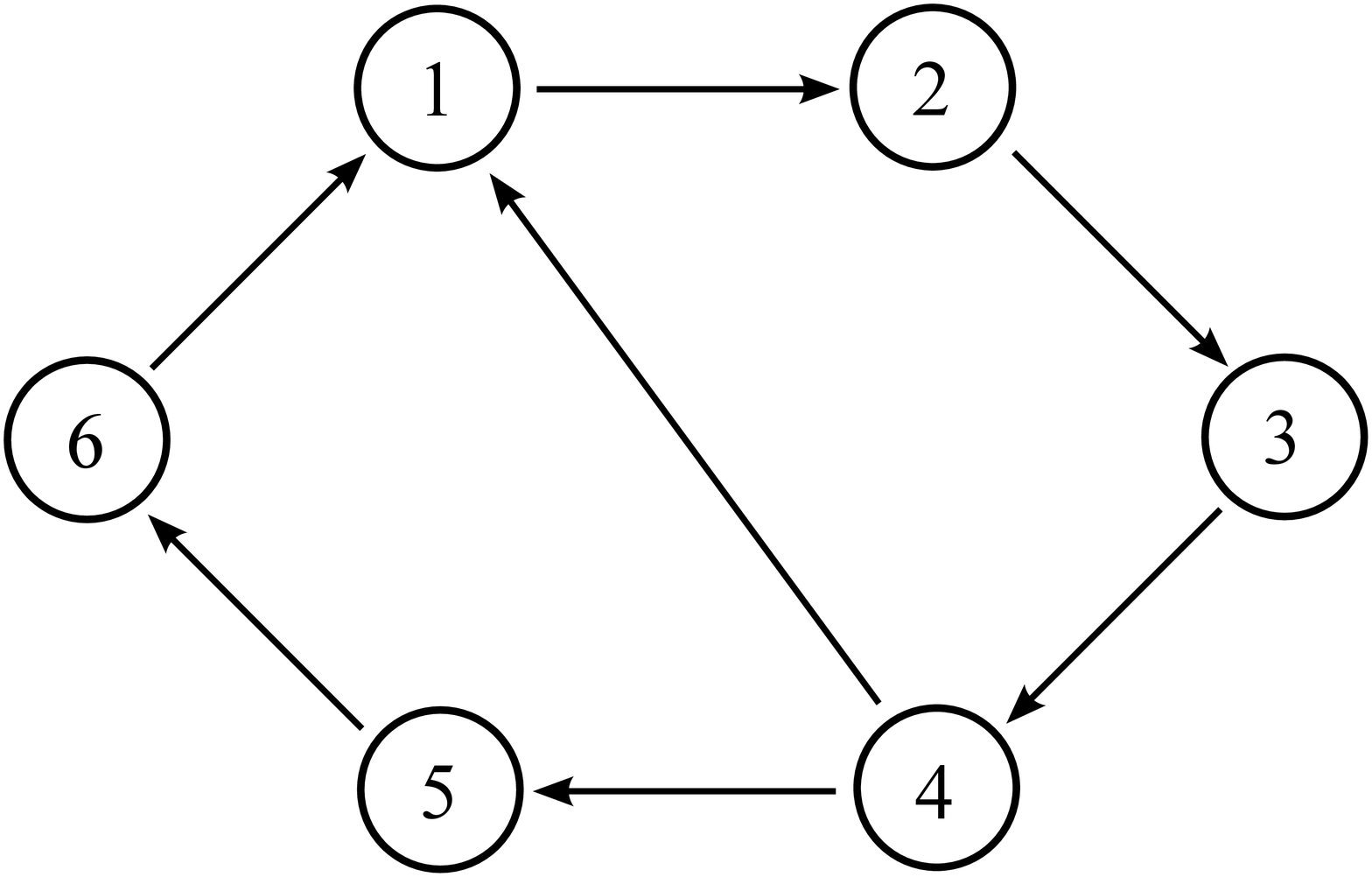}}\\
a) & b) \\[20pt]
\scalebox{0.16}{\includegraphics{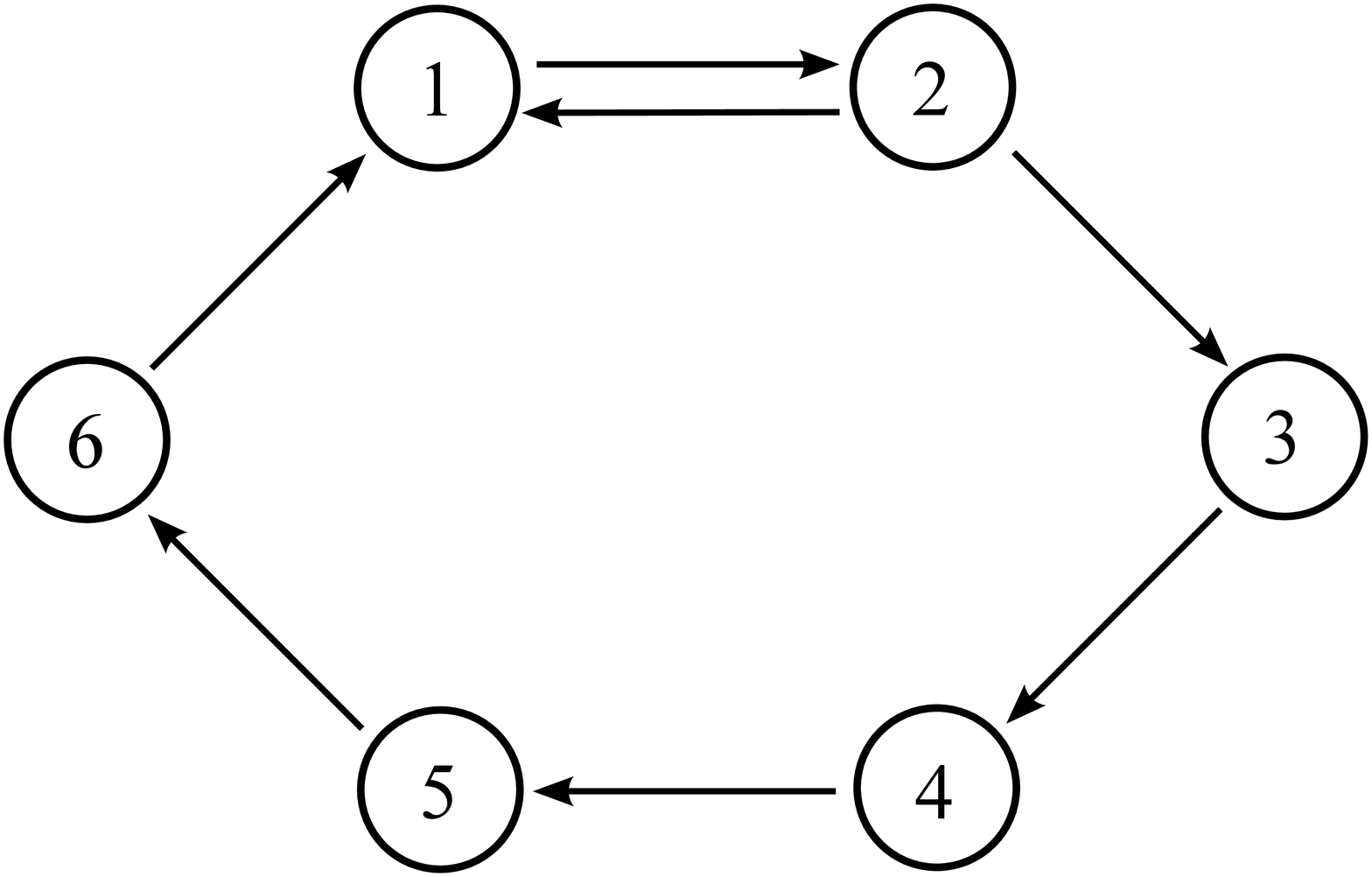}} & \scalebox{0.16}{\includegraphics{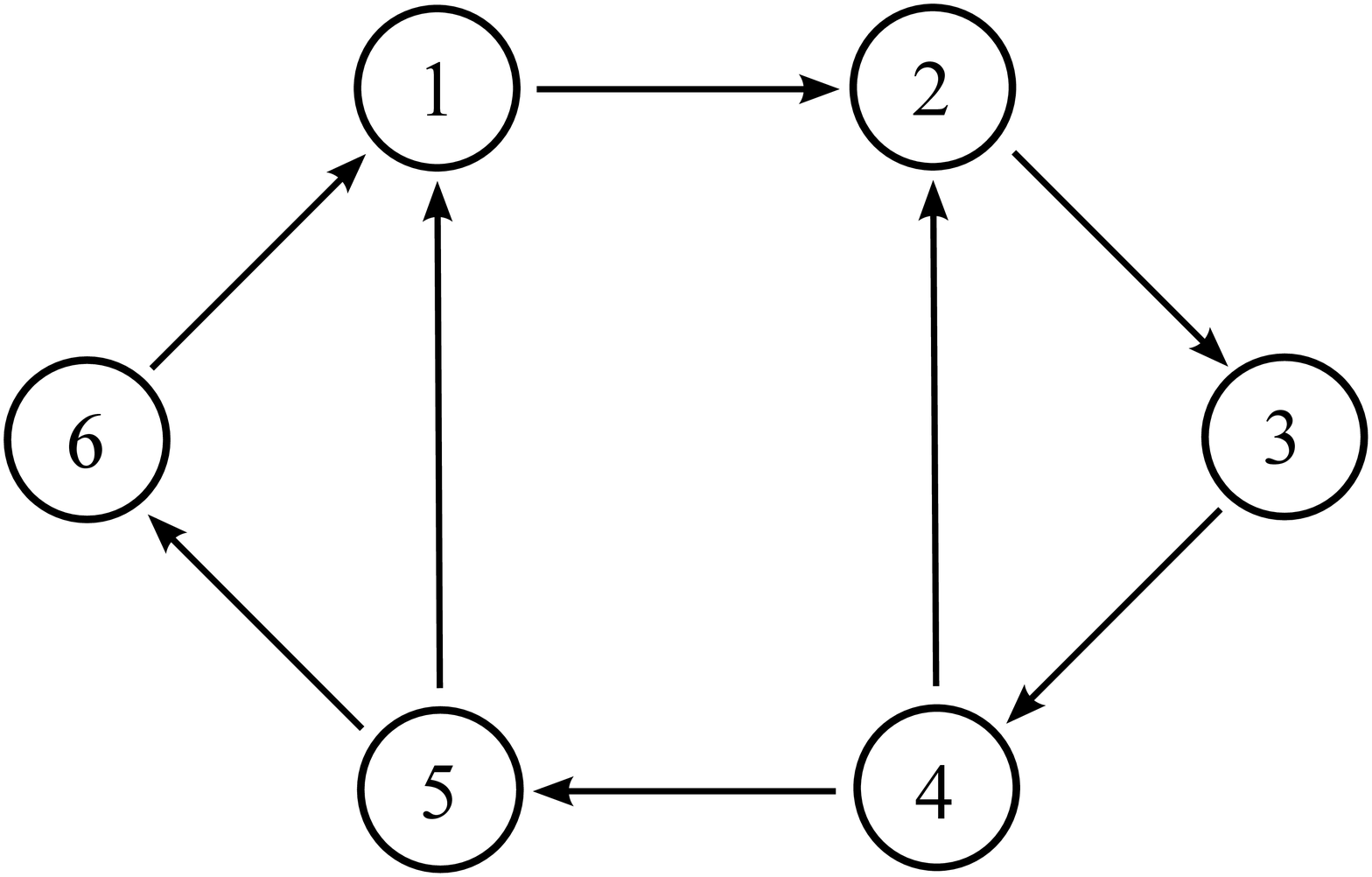}}\\
c) & d)
\end{tabular}
\caption{Network Graphs similar to $G^{\sf prio}$}
\label{fig:morePrioArbiters}
\end{figure}

\begin{exa}
Figure~\ref{fig:morePrioArbiters} shows several network graphs that
are similar to $G^{\sf prio}$ from
Example~\ref{ex:PrioArbiter}. Graphs a) and b) have the same
$2$-topology, while c) and d) have not: the connection topology for c)
and $R=\{1,2\}$ is not in $TC_2(G^{\sf prio})$ (it has direct back and
forth connections between $i$ and $j$), and the same holds for the
topology of d) and $R=\{1,4\}$ (which is similar to 
Figure~\ref{fig:2Topo} b), but has back and forth connections from $i$ to
both hubs). 

Note that, if we allow quantifier alternations, even a) and b) do not
agree on all $2$-indexed specifications: assuming that we have an
implementation $\mT$ such that we can directly observe that a given
process has the token, an example for a $2$-indexed specification that
does not hold in all graphs with the $2$-topology $CT_2(G^{\sf prio})$
is 

\[ \forall i.~ \exists j.~ \left( \begin{array}{ll} 
	& (\token_i \impl \token_i \weakuntil \token_j) \land (\neg \token_j \impl \neg \token_j \weakuntil \token_i)\\
\lor & (\token_j \impl \token_j \weakuntil \token_i) \land (\neg \token_i \impl \neg \token_i \weakuntil \token_j)
\end{array} \right). \]
This formula is valid in the graphs from Figure~\ref{fig:PrioArbiter} and Figure~\ref{fig:morePrioArbiters} b), but not in  Figure~\ref{fig:morePrioArbiters} a): for $i=8$ there is no $j$ such that the formula holds.
\end{exa}

\subsection{Parameterized Synthesis in Token Networks}
\label{sec:synthesis-networks}
We can lift these verification results to the synthesis of parameterized token-passing networks, 
such that the resulting implementation will ensure a given 
$k$-indexed specification in all networks with a given reduction $(CT_k,B(g^1,\ldots,g^m))$.

To synthesize process implementations in token networks, we again adapt the bounded synthesis approach introduced in Section~\ref{sec:bounded}. We first show how to encode the isomorphic synthesis problem for a single connection topology $CT \in CT_k$. To this end, we use the general modifications mentioned in Section~\ref{sec:adapted} for isomorphic processes and small implementations, as well as the binary representation of scheduling variables. Additionally, we use the following modifications:
\begin{itemize}
\item Hub nodes are not synthesized, but have a fixed implementation with three states: one where they wait for the token, one they enter upon receiving the token, and another one where $\send$ is active, which is entered when they are scheduled again after receiving the token.
\item The special features of token-passing networks are encoded
  similarly to token rings. The main difference is that we cannot talk
  about process $i+1$ anymore, since successors with respect to token
  passing are not unique in general token networks. Instead, the corresponding constraints talk about all processes which have a connection to the process that is sending the token. In the following, $(i,j)\in CT$ means there is a direct connection between processes $i$ and $j$ in $CT$, and $P^-$ stands for non-environment processes (in both cases including hubs):

\begin{align*}
  \bigwedge_{i \in P^-}~\bigwedge_{t \in T}~ \hspace{48pt}&
	\neg \token(d_i(t)) \impl \neg \send(d_i(t))\\[5pt]	
  \bigwedge_{(i,j)\in CT}~\bigwedge_{t \in T}~\bigwedge_{I \in \mathcal{P}(O_{env})}~&
	\send(d_i(t)) \land \sched_{j}(I) \impl \neg \token(d_i(\delta(t,I)))\\[5pt]
  \bigwedge_{(i,j) \in CT}~\bigwedge_{t \in T}~\bigwedge_{I \in \mathcal{P}(O_{env})}~&
	\send(d_{i}(t)) \land \sched_j(I) \impl \token(d_j(\delta(t,I)))\\[5pt]	
\end{align*}
\begin{align*}
  \bigwedge_{(i,j) \in CT}~\bigwedge_{t \in T}~\bigwedge_{I \in \mathcal{P}(O_{env})}~&
	\token(d_i(t)) \impl (\send(d_i(t)) \land \sched_j(I)) \lor \token(d_i(\delta(t,I)))\\[5pt]	
	\bigwedge_{(i,j) \in CT}~\bigwedge_{t \in T}~\bigwedge_{I \in \mathcal{P}(O_{env})}~&
	\neg \token(d_j(t)) \land \neg (\send(d_i(t)) \land \sched_j(I))\impl \neg \token(d_j(\delta(t,I))) \hspace{-36pt}
\end{align*}
\item Similarly, the restriction of state changes to the scheduled process, with exception of token-passing steps, needs to be modified to account for connections in the topology:

\[\bigwedge_{(i,j) \in CT}~\bigwedge_{t \in T}~\bigwedge_{I \in \mathcal{P}(O_{env})}~ 
	\neg \sched_i(I) \land \neg (\send(d_i(t)) \land \sched_j(I))
	\impl d_i(\delta(t,I)) = d_i(t)\hspace{-36pt} \]

\item To ensure that we obtain isomorphic constraints, we pick one process from $CT$ (denoted by $1$ below) and modify the corresponding constraints from Section~\ref{sec:adapted} to the following, where $P^*$ stands for $P^-$ without hubs and without $1$:

\[
  \bigwedge_{i \in P^*}~\bigwedge_{t,t' \in T}~\bigwedge_{I,I' \in \mathcal{P}(O_{env})}~
	\left( \begin{array}{l} 
					d_1(t) = d_i(t') \land \sched_1(I) \land \sched_i(I') \land I \cap I_1 = I' \cap I_i\\[5pt]
					\impl d_1(\delta(t,I)) = d_i(\delta(t',I')) 
					\end{array} \right) \hspace{-38pt} \]
\[	\bigwedge_{(i,j) \in CT, i \in P^*}~\bigwedge_{ (1,l) \in CT }~\bigwedge_{t,t' \in T}~\bigwedge_{I,I' \in \mathcal{P}(O_{env})}~
  \left( \begin{array}{l}
					d_1(t) = d_i(t') \land \send(d_1(t)) \land \send(d_i(t')) \\[5pt]
					\land~ \sched_l(I) \land \sched_j(I') \land I \cap I_1 = I' \cap I_i \\[5pt]
					\impl d_1(\delta(t,I)) = d_i(\delta(t',I'))
				 \end{array} \right) \hspace{-16pt} \]
	
\item As for token rings, we ensure that the synthesized process
  implementation will always eventually release the token under the
  assumption of fair scheduling by adding the same constraint:
\[ \bigwedge_{i \in P^-}~ {\sf fair\_scheduling} \impl \always (\token_i \impl \eventually \send_i). \]
However, since the environment decides on the connection used for
token-passing in case of multiple connections, this property does not
imply fair token passing, even under the assumption of fair
scheduling. Thus, instead of ${\sf fair\_scheduling}$, we directly add  
fair token passing (${\sf fair\_token} \equiv \forall i.~ \always \eventually \token_i$) as an
environment assumption to all liveness guarantees of the
system.\footnote{Note that this differs from what we claimed
  before~\cite{JacobsBloem12}, due to a misunderstanding of how token
  passing was defined by Clarke et al.~\cite{Clarke04c}.} 
\item Finally, since we want to synthesize processes that potentially have to satisfy constraints from several $CT$ at the same time, we do not use a bound on the size of the overall system implementation, but instead on the process implementation. This can be achieved e.g. by letting the $d_i$ map into $\{1,\ldots,n\}$, and choosing the size of the overall system as $n^{\card{P^*}} \cdot 3^h$, where $h$ is the number of hubs in $CT$.
\end{itemize}

\noindent With these modifications, we can use the bounded synthesis approach to encode the isomorphic synthesis problem for a $k$-indexed formula $\spec(\overline{i})$ and a connection topology $CT$ into a sequence of set of SMT constraints $SMT(CT,n)$, where $n$ is the bound on the size of process implementations.

In order to extend this to parameterized synthesis for all networks
with the same reduction $(CT_k, B(g_1,\ldots,g_m))$, we encode the
isomorphic synthesis problem $A_{CT}, \mT \models \spec(\overline{i})$
for all $CT \in CT_k$, using for all $CT$ the same function symbols
for outputs and transition function of the process to be
synthesized. If the original specification universally quantifies over
all indices, the SMT constraint we are trying to solve is
$\bigwedge_{CT \in CT_k} SMT(CT,n)$, for increasing $n$. In general,
we need to solve $B(SMT(CT^1,n),\ldots,SMT(CT^m,n))$. As
before, if for a given bound no solution exists, we increase $n$ until
an implementation is found. 

Note that, in contrast to verification, we cannot solve the problems
$SMT(CT^i,n)$ independently: we want to obtain an implementation that satisfies all
of these constraints (or a Boolean combination of the constraints),
and thus have to consider the combination of the constraint systems. 

\begin{thm}
Let $\mG$ be a class of network graphs, $\pspec(\overline{i})$ a
$k$-indexed parametric specification, and $(CT_k,B(g^1,\ldots,g^m))$ a
reduction for $\mG$ and $\pspec(\overline{i})$. If
$\pspec(\overline{i})$ is realizable in all token-passing networks
based on graphs in $\mG$, then the adapted bounded synthesis
procedure will eventually find an implementation that satisfies
$\pspec(\overline{i})$ in all these token-passing networks.
\end{thm}


\subsubsection*{Symmetry Reduction}
Since the number of different topologies in $CT_k$ can be very big, and to a large extent consists of symmetric variants, we propose the following optimization: if the specification $\pspec(\overline{i})$ is symmetric, i.e. we have $\pspec(\overline{i}) \Leftrightarrow \pspec(\overline{j})$ for any permutation $\overline{j}$ of $\overline{i}$, then we do not need to consider symmetric variants. For specifications with many variables, we can make this even more fine-grained: if we have $\pspec(\ldots,i,j,\ldots) \Leftrightarrow \pspec(\ldots,j,i,\ldots)$, then for each pair of connection topologies such that one can be obtained from the other by switching positions of $i$ and $j$, we only need to consider one of them.

This optimization can be used both for model checking and synthesis of implementations in token-passing networks.

\begin{exa}
When considering symmetric specifications in the prioritized
token-ring architecture, we only check the $6$ topologies depicted in
Figure~\ref{fig:2Topo}, and not the $5$ additional symmetric
variants. This simplifies the SMT encoding of the specification, and
the resulting constraint will be a conjunction of 
the constraints for $6$ topologies, instead of $11$. 
\end{exa}

\section{Synthesizing a Parameterized Arbiter}
\label{sec:implementation}

In this section, we show how parameterized synthesis can be used
to obtain process implementations for token ring architectures and
general token-passing networks, exemplified by prioritized token
rings. Our example is a parameterized arbiter with the following
specification $\forall i,j.~\spec(i,j)$: 

\[ \begin{array}{ll}
  \forall i \neq j.~ & \always \neg ( g_i \land g_j ) \\
  \forall i.~ & \always (r_i \impl \eventually g_i).
  \end{array} 
\]

\medskip \noindent
Every process $i$ has an input $r_i$ for requests from
the environment and an output $g_i$ to grant requests. We
want grants of all processes to be mutually exclusive, and every request to be
eventually followed by a grant. 

\subsection{Token Rings.} 
The arbiter specification satisfies case c) in
Theorem~\ref{thm:reduction}, i.e., a ring of size 4 is sufficient to synthesize
implementations that satisfy the specification for rings of any size.

According to the adapted bounded synthesis approach from
Section~\ref{sec:adapted}, we need to add the token fairness requirement, and add
the fair scheduling assumption to all liveness constraints. This results in the
extended specification $\pspec$:

\[ \begin{array}{ll}
  \forall i \neq j.~ & \always \neg ( g_i \land g_j ) \\
  \forall {i}. & {\sf fair\_scheduling} \impl \always (r_i \impl \eventually g_i)\\
  \forall {i}. & {\sf fair\_scheduling} \impl \always (\token_i \impl \eventually \send_i).
  \end{array} 
\]

\medskip \noindent
We translate the specification into a universal co-\Buchi automaton, shown for 2
processes in Figure~\ref{fig:automaton}. The $\bot$-state is universally rejecting, i.e. any 
trace that visits it is rejected by the automaton. This automaton
translates to a set of first-order constraints for the annotations of an LTS
implementing $\pspec$, a part of which is shown in
Figure~\ref{fig:constraints} (only constraints for states $0,1,3,5$ of the
automaton are shown). These constraints, together with general
constraints for asynchronous systems, isomorphic processes,
token rings, and size bounds, are handed to Z3~\cite{Moura08}. For correctly chosen bounds
($\card{T_A} \leq 4$ and $\card{T_p} \leq 2$), we obtain a model of the process
implementation in $\sim$5.5 seconds (on an Intel Core i5 CPU @ 2.60GHz).

\begin{figure}
\begin{pspicture}(-4,-1.5)(7,9.9)
\psset{arrows=->,nodesep=5pt}
\put(0.5,4){\rnode{emt}{}}
\cnodeput(2,4){0}{$0$}
\cnodeput(-0.5,2){1}{$1$}
\put(-0.5,2){\Cnode[radius=0.35cm]{a}}
\put(1.87,2){\rnode{err}{$\bot$}}
\cnodeput(4.5,2){2}{$2$}
\put(4.5,2){\Cnode[radius=0.35cm]{a}}
\cnodeput(-2,0){3}{$3$}
\cnodeput(1,0){5}{$5$}
\cnodeput(6,0){4}{$4$}
\cnodeput(3,0){6}{$6$}

\ncline{0}{1}\tlput{$r_1\overline{g_1}$}
\ncline{0}{err}\trput{$g_1g_2$}
\ncline{0}{2}\trput{$r_2\overline{g_2}$}
\nccircle[angle=0,nodesep=2pt]{0}{0.25cm}\taput{$*$}
\ncline{emt}{0}
\ncarc[arrows=<-]{3}{1}\tlput{$s_1\overline{g_1}$}
\nccircle[angle=180,nodesep=2pt]{3}{0.25cm}\tbput{$s_1\overline{g_1}$}
\ncarc[arrows=<-]{5}{3}\tbput{$s_2\overline{g_1}$}
\nccircle[angle=180,nodesep=2pt]{5}{0.25cm}\tbput{$s_2\overline{g_1}$}
\ncarc[arrows=<-]{1}{5}\trput{$s_1\overline{g_1}$}
\ncarc[arrows=<-]{5}{1}\tlput{$s_2\overline{g_1}$}

\ncarc{2}{4}\trput{$s_2\overline{g_2}$}
\ncarc{2}{6}\trput{$s_1\overline{g_2}$}
\ncarc{4}{6}\tbput{$s_1\overline{g_2}$}
\nccircle[arrows=<-,angle=180,nodesep=2pt]{4}{0.25cm}\tbput{$s_2\overline{g_2}$}
\nccircle[arrows=<-,angle=180,nodesep=2pt]{6}{0.25cm}\tbput{$s_1\overline{g_2}$}
\ncarc{6}{2}\tlput{$s_2\overline{g_2}$}

\cnodeput(-1,6){7}{$7$}
\put(-1,6){\Cnode[radius=0.35cm]{a}}
\cnodeput(5,6){8}{$8$}
\put(5,6){\Cnode[radius=0.35cm]{a}}
\cnodeput(-3,8){9}{$9$}
\cnodeput(0.5,8){11}{$11$}
\cnodeput(7,8){10}{$10$}
\cnodeput(3.5,8){12}{$12$}

\ncline{0}{7}\tlput{$\token_1\overline{\send_1}$}
\ncline{0}{8}\trput{$\token_2\overline{\send_2}$}
\ncarc{7}{9}\tlput{$s_1\overline{\send_1}$}
\nccircle[arrows=<-,angle=0,nodesep=2pt]{9}{0.25cm}\taput{$s_1\overline{\send_1}$}
\ncarc{9}{11}\taput{$s_2\overline{\send_1}$}
\nccircle[arrows=<-,angle=0,nodesep=2pt]{11}{0.25cm}\taput{$s_2\overline{\send_1}$}
\ncarc{11}{7}\trput{$s_1\overline{\send_1}$}
\ncarc{7}{11}\tlput{$s_2\overline{\send_1}$}

\ncarc[arrows=<-]{10}{8}\trput{$s_2\overline{\send_2}$}
\ncarc[arrows=<-]{12}{8}\trput{$s_1\overline{\send_2}$}
\ncarc[arrows=<-]{12}{10}\taput{$s_1\overline{\send_2}$}
\nccircle[angle=0,nodesep=2pt]{10}{0.25cm}\taput{$s_2\overline{\send_2}$}
\nccircle[angle=0,nodesep=2pt]{12}{0.25cm}\taput{$s_1\overline{\send_2}$}
\ncarc[arrows=<-]{8}{12}\tlput{$s_2\overline{\send_2}$}

\end{pspicture}
\caption{Universal co-\Buchi automaton for specification $\pspec$}
\label{fig:automaton}
\end{figure}

\begin{figure}
\[
\begin{array}{ll}
& \lambda^\bbB_0 (0)\\
& \token(d_1(0)) \land \neg \token(d_2(0))\\
\bigwedge_{t \in T}~\bigwedge_{I \in \mathcal{P}(O_{env})}~& \lambda^\bbB_0(t) \impl \lambda^\bbB_0(\delta(t,I)) \land
\lambda^\#_0(\delta(t,I)) \geq \lambda^\#_0(t)\\
\bigwedge_{t \in T} & \lambda^\bbB_0(t) \impl \neg ( g(d_1(t)) \land
g(d_2(t)))\\
\bigwedge_{t \in T} \bigwedge_{I \in \mathcal{P}(O_{env})}~& \lambda^\bbB_0(t) \land \sched_1(I) \land r_1
\in I \land \neg g(d_1(t)) \impl \lambda^\bbB_1(t) \land \lambda^\#_1(\delta(t,I)) > \lambda^\#_0(t)\\
\bigwedge_{t \in T} \bigwedge_{I \in \mathcal{P}(O_{env})} & \lambda^\bbB_1(t) \land \neg \sched_2(I)
\land \neg g(d_1(t)) \impl \lambda^\bbB_3(t) \land
\lambda^\#_3(\delta(t,I)) \geq \lambda^\#_1(t)\\
\bigwedge_{t \in T} \bigwedge_{I \in \mathcal{P}(O_{env})} & \lambda^\bbB_1(t) \land \sched_2(I)
\land \neg g(d_1(t)) \impl \lambda^\bbB_5(t) \land
\lambda^\#_5(\delta(t,I)) \geq \lambda^\#_1(t)\\
\bigwedge_{t \in T} \bigwedge_{I \in \mathcal{P}(O_{env})} & \lambda^\bbB_3(t) \land \neg \sched_2(I)
\land \neg g(d_1(t)) \impl \lambda^\bbB_3(t) \land \lambda^\#_3(\delta(t,I)) \geq \lambda^\#_3(t)\\
\bigwedge_{t \in T} \bigwedge_{I \in \mathcal{P}(O_{env})} & \lambda^\bbB_5(t) \land \neg \sched_1(I)
\land \neg g(d_1(t)) \impl \lambda^\bbB_5(t) \land
\lambda^\#_5(\delta(t,I)) \geq \lambda^\#_5(t)\\
\bigwedge_{t \in T} \bigwedge_{I \in \mathcal{P}(O_{env})} & \lambda^\bbB_5(t) \land \sched_1(I)
\land \neg g(d_1(t)) \impl \lambda^\bbB_1(t) \land \lambda^\#_1(\delta(t,I)) >
\lambda^\#_5(t)\\
\hfill \ldots \hfill & \hfill \ldots \hfill
\end{array}
\]
\caption{Constraints that are equivalent to realizability of $\pspec$}
\label{fig:constraints}
\end{figure}

The solution is very simple: every process needs only 2 states, with
$\send_i$ and $g_i$ signals high if and only if the process has the token. In the parallel
composition of 4 such processes, only 4 global states are
reachable.
Theorem~\ref{cor:synthesis:reduction} guarantees that with 
this process implementation, $\pspec$ will be satisfied for any instance of the
architecture. Figure~\ref{fig:arbiter_ring_one} depicts the LTS for one
process, and Fig.~\ref{fig:arbiter_ring_parallel} the parallel
composition of 4 processes in a ring. 

\begin{figure}
\begin{subfigure}[b]{0.33\linewidth}
    \centering
      \begin{tikzpicture}[->,
node distance = 2.5cm, 
auto, 
thick, 
inner sep=.1cm,
bend angle=35]
    \tikzset{every state/.style={rectangle,rounded corners,minimum size=2em}}
    \tikzset{every edge/.append style={font=\small, right}}
    \tikzset{box state/.style={draw,rectangle,rounded corners,inner sep=.1cm}}
    
    \node[state] (1) {$\neg \token_i~\neg g_i~\neg \send_i$};
    \node[state] (2) [below of=1] {$\token_i~g_i~\send_i$};
    
    \path (1) edge[loop above] node[anchor=south,above] {$\neg \send_{i-1}$} (1);
    \path (1)  edge[bend left] node[anchor=west,right,xshift=2pt] {$\send_{i-1}$} (2);
    \path (2)  edge[bend left] node[anchor=east,left,yshift=-2pt] {$*$} (1);
    
    
    
    
    
    
\end{tikzpicture}
      \caption{Process implementation}
      \label{fig:arbiter_ring_one}
 \end{subfigure}
 \begin{subfigure}[b]{0.63\textwidth}
    \centering
      \begin{tikzpicture}[->,
node distance = 2.4cm, 
auto, 
thick, 
inner sep=.1cm,
bend angle=30]
    \tikzset{every state/.style={rectangle,rounded corners,minimum size=2em}}
    \tikzset{every edge/.append style={font=\small, right}}
    \tikzset{box state/.style={draw,rectangle,rounded corners,inner sep=.1cm}}
    
    \node[state] (1) [text width=4.1cm,align=right] {$\token_1~g_1~\send_1$~\\$\forall i{\neq}1:\neg \token_i~\neg g_i~\neg \send_i$~};
    \node[state] (2) [below right of=1,text width=4.1cm,align=right,xshift=1cm] {$\token_2~g_2~\send_2$~\\$\forall i{\neq}2:\neg \token_i~\neg g_i~\neg \send_i$~};
    \node[state] (4) [below left of=1,text width=4.1cm,align=right,xshift=-1cm] {$\token_4~g_4~\send_4$~\\$\forall i{\neq}4:\neg \token_i~\neg g_i~\neg \send_i$~};
    \node[state] (3) [below right of=4,text width=4.1cm,align=right,xshift=1cm] {$\token_3~g_3~\send_3$~\\$\forall i{\neq}3:\neg \token_i~\neg g_i~\neg \send_i$~};
    
    \path (1)  edge[bend left] node[anchor=south,above,yshift=2pt] {} (2);
    \path (2)  edge[bend left] node[anchor=north,below,xshift=-3pt,yshift=-3pt] {} (3);
    \path (3)  edge[bend left] node[anchor=north,below,xshift=-3pt,yshift=-3pt] {} (4);
    \path (4)  edge[bend left] node[anchor=north,below,xshift=-3pt,yshift=-3pt] {} (1);
    
    
    
    
    
    
\end{tikzpicture}
      \caption{Parallel composition in ring of $4$}
      \label{fig:arbiter_ring_parallel}
 \end{subfigure}

\caption{Parameterized arbiter implementations}
\end{figure}

Note that synthesis is easy in this case because we can restrict it to a
small ring of 4 processes, and have a rather simple specification. For 5
processes (and $\card{T_A} \leq 5$), Z3 already needs $\sim$100
seconds to solve the resulting constraints.

\subsection{Prioritized Token Rings.} 

Now, we consider the arbiter specification $\forall i,j.~\spec(i,j)$
from above, and the $2$-topology $CT_2(G^{\sf prio})$ for prioritized
token rings, as given in Figure~\ref{fig:2Topo} (modulo symmetric
variants). We want to find an implementation $\mT$ such that 
$A_{G^{\sf prio}}, \mT \models \forall i,j.~ \spec(i,j)$, under the
assumption of fair token passing.

According to the synthesis approach from
Section~\ref{sec:synthesis-networks}, we add fair token passing as an
environment assumption to the liveness constraint of the arbiter, and
add a constraint that ensures that every process must eventually
release the token if scheduling is fair. This results in the following
specification $\pspec'$:

\[ \begin{array}{ll}
  \forall i \neq j.~ & \always \neg ( g_i \land g_j ) \\
  \forall {i}. & {\sf fair\_token} \impl \always (r_i \impl \eventually g_i)\\
  \forall {i}. & {\sf fair\_scheduling} \impl \always (\token_i \impl \eventually \send_i).
  \end{array} 
\]
As before, the specification is translated into a universal
co-B\"uchi automaton. Then, for all $11$ 
topologies $CT \in CT_2(G^{\sf prio})$, we translate this automaton
into a set of constraints $SMT(CT,n)$, using the same function
symbols for transition and output functions of the synthesized
process in each set of constraints. Then, we solve the conjunction of
all these constraints for 
increasing $n$. 

Again, a solution is very simple, and in fact the
same process implementation that satisfies this specification in rings
(see Figure~\ref{fig:arbiter_ring_one}) is synthesized in this
case.\footnote{The fact that the same implementation works in this
  case may seem counterintuitive. The reason is that we only
  consider executions with fair scheduling, and the
  scheduler is part of the environment. Thus, the process
  implementation only needs to guarantee that it only gives a grant if
  it has the token, and that it will eventually release the token.}
For the correct size bound of $2$, Z3 needs ${\sim}27$ seconds to solve
the resulting constraints.

Since the arbiter specification is symmetric ($\spec(i,j) \Leftrightarrow
\spec(j,i)$), we can use the symmetry
reduction technique mentioned at the end of
Section~\ref{sec:network}. That is, we only need to consider the $6$
topologies depicted in Figure~\ref{fig:2Topo}, and not the $5$
additional symmetric variants. The resulting SMT constraints have a size of ${\sim}
2.8$MB instead of ${\sim}5.1$MB, and are solved by Z3 in ${\sim} 5$
seconds instead of ${\sim} 27$.

\subsubsection*{The Parameterized Synthesis Tool \textsc{Party}.} 
The experiments presented above are rather restricted because part of
the translation of specifications into SMT constraints was done
manually. Khalimov, Jacobs and Bloem~\cite{Khalimov13a} have since
developed a fully automatic implementation of the approach for token rings, and
compared the time required for parameterized synthesis for several
different benchmarks and combinations of optimizations. In particular,
they show that significant increases in synthesis time, similar to
those for increasing number of components, can also be observed if we
consider more complex specifications (in a ring of the same size).

Based on the synthesis approach presented in this paper,
\textsc{Party}~\cite{Khalimov13a} implements additional optimizations
and extensions due to Khalimov, Jacobs and Bloem~\cite{Khalimov13}. The
tool accepts a specification in indexed LTL, in a language derived
from that of (monolithic) synthesis tool Acacia+~\cite{Bohy12}. Based 
on the syntactical form of the specification, it
automatically determines the valid cutoff for an implementation in a
token-ring architecture, and applies our synthesis method with
suitable optimizations. 

\section{A Framework for Parameterized Synthesis}

Our approach for reduction of parameterized synthesis to isomorphic
synthesis is not limited to token-passing systems. The methods presented here can be 
seen as the basis of a framework that lifts certain classes of
algorithms for the verification of parameterized systems to 
(semi-)algorithms for their synthesis.

There is a vast body of work on the verification of
parameterized systems, much of it going beyond token-passing
systems. In the following, we consider the 
problem of lifting these results to parameterized synthesis. The methods 
described in this paper extend more or less directly to results
that provide a cutoff that reduces
the parameterized model checking problem to 
an equivalent set of finite-state model checking problems. In addition
to these, we survey other methods for parameterized
verification, and how they might be lifted to parameterized
synthesis.

\subsection{Methods Based on Cutoffs.}
The literature on parameterized model checking contains many
results that prove a cutoff for the given class of systems 
and specifications, making the parameterized verification problem
decidable. We give an incomplete overview: 
\begin{itemize}
\item  German and Sistla~\cite{German92} provide cutoffs for
  $1$-indexed properties in architectures with pairwise communication
  (synchronization) in a clique. Intended application areas are
  resource allocation algorithms and network protocols.
\item Emerson and Kahlon~\cite{Emerson00} provide cutoffs for systems
  where transitions of any component are guarded with conjunctive or
disjunctive statements about the states of the other processes,
effectively constituting a clique structure with a limited form of
shared variables. 
Intended applications are cache coherence protocols and
readers-writers problems.
\item Furthermore, Emerson and Kahlon~\cite{Emerson03} provide cutoffs for 
  (initialized) broadcast protocols that can be used for proving
  cache coherency.
\item Additionally, Emerson and Kahlon~\cite{Emerson04} provide
  cutoffs for bi-directional rings with multi-valued tokens and some
  additional restrictions.  Intended applications are leader election
  algorithms, as well as resource allocation algorithms.
\item Kahlon et al.~\cite{Kahlon05} show that also for threads
  communicating via locks, there are cutoffs for certain cases. Like 
	the guarded transitions of Emerson and Kahlon~\cite{Emerson00}, 
	this can be seen as a limited form of shared variables.
\item Bouajjani et al.~\cite{Bouajjani08} consider resource management
  systems based on (prioritized) FIFO queues, and provide cutoffs for
  several cases.
\item Aminof et al.~\cite{AJKR14} extend and unify the results
  for token rings and general token-passing networks considered in
  this paper. In particular, they provide concrete cutoffs for
  processes arranged in rings, cliques, or stars.
\end{itemize}
In principle, any verification result that provides a cutoff can be used 
to obtain a semi-decision procedure for the parameterized synthesis
problem. We distinguish three cases:

\subsubsection*{Static Structure-Independent Cutoffs.}
For many of the results mentioned
above~\cite{Emerson00,Emerson04,Kahlon05,Bouajjani08,AJKR14}, the cutoff 
depends only on the architecture and the specification, but not on the
(structure of the) implementation. In this case,
the approach is directly analogous to what we described for
token-passing systems: 
\begin{enumerate} 
\item determine suitable cutoff, based on architecture and specification,
\item encode synthesis problem into SMT constraints for bounded
  synthesis\\ (including architecture-specific encoding), and
\item for increasing $n$, until implementation is found:\\ solve
  bounded synthesis problem by solving SMT constraints with size
  bound $n$.
\end{enumerate}
In this case, the only limitation is the ability to (efficiently)
encode the features of the class of systems in decidable
first-order constraints. This should be possible for all of 
the results mentioned above.

\subsubsection*{Static Structure-Dependent Cutoffs.}
For some of the results~\cite{German92,Emerson00}, the cutoff also
depends on the size (i.e., the number of states) of the
implementation. In this case, all three steps need to be repeated
whenever we want to check if an implementation for a given size
exists:

For increasing $n$, until implementation is found:
\begin{enumerate}
\item determine suitable cutoff, based on architecture,
  specification, and $n$,
\item encode synthesis problem into SMT constraints for bounded
  synthesis\\ (including architecture-specific encoding), and
\item solve bounded synthesis problem by solving SMT
  constraints for size $n$.
\end{enumerate}

\medskip
\noindent
Additionally, there are results~\cite{Bouajjani08} where the cutoff not
only depends on the size of the implementation, but some other
properties, e.g., the number of transitions. For these, the synthesis
approach has to be refined again, enumerating models with
increasing number of transitions together with their suitable
cutoff. Similar approaches can be used for cutoffs that depend on
other properties (e.g., the diameter) of the implementation, but may
be much more difficult to implement efficiently than the comparably
simple approaches above.

\subsubsection*{Dynamic Cutoffs.}
There are also approaches that detect a cutoff for a given system
implementation dynamically~\cite{Hanna09,Kaiser10}. That is, the cutoff
is not determined by syntactic properties of architecture,
specification, or implementation. These are less suited for our
framework: in order to integrate them with our approach, cutoff
detection would have to be interleaved with generation of candidate
implementations, making it hard to devise a complete
synthesis approach. Finding out how this can be done constitutes a
separate direction of research.

\subsection{Other Methods.}
There are several other results for the verification of systems with
an arbitrary number of components. For these, it is less clear how to
lift results from the verification of parameterized systems to their
synthesis. Again, we distinguish three cases:

\subsubsection*{Induction-Based Approaches.} 
These approaches reduce the parameterized verification
  problem to the problem of finding an inductive
  \emph{network invariant}~\cite{Kurshan95}. However, they are usually
  not guaranteed to work for a fixed class of systems, and the
  invariant must be found manually for the given system under
  consideration. Clarke et al.~\cite{Clarke97} introduced a method
  that partly automates the construction of network
  invariants. Finally, the \emph{invisible invariants}
  approach~\cite{Pnueli01,Zuck04} can be seen as a  combination of
  network invariants with automatic detection of cutoffs (that can
  directly depend on the invariant to be proved).

Since network invariants not only depend on the specification, but
also on the implementation, a parameterized synthesis approach would
have a feedback loop between synthesis and invariant generation,
similar to the case of dynamic cutoffs.

\subsubsection*{Abstraction-Based Methods.}
These methods construct a finite-state
  abstraction of a parameterized system, and have been pioneered by
  \emph{counter abstraction}~\cite{Pnueli02,Zuck04}. An interesting
  extension of this approach combines counter abstraction with
  \emph{environment abstraction}, which centers on one process, and
  abstracts the behavior of all other
  processes~\cite{Clarke06,Clarke08}. 

A way to integrate such results into a parameterized synthesis method
would be to identify conditions on distributed systems that guarantee
applicability of, e.g., counter-abstraction, and then apply synthesis
modulo these conditions. Consequently, synthesis will 
find an implementation if and only if there exists one that satisfies 
these conditions. 

\subsubsection*{Regular Model Checking.}
Regular model checking~\cite{Bouajjani00} is an approach for the
verification of infinite-state systems that can also be used for
parameterized verification. It is based on the idea that the state of
a system can be expressed as a regular expression, and transitions
given as finite-state transducers.
Overviews of some of the methods based on regular model
checking can be found in the work of 
Vojnar~\cite{Vojnar07} and Abdulla~\cite{Abdulla12}.

The question how to integrate regular model checking into
parameterized synthesis is wide open. One option would be to
consider approaches that compute (approximations of) the reachable
states in regular model checking, and try to extend them to a
game-based synthesis approach.

\section{Conclusions and Future Work}
\label{sec:conclusion}
We have stated the problem of parameterized realizability and
parameterized synthesis: whether and how a parameterized specification
can be turned into a simple recipe for constructing a parameterized
system. The realizability problem asks whether a parameterized
specification can be implemented for any number of processes, i.e.,
whether the specification is correct. Our procedure for parameterized
synthesis yields a process implementation that can be replicated to
obtain a correct system of arbitrary size, thus avoiding the steeply
rising need for resources associated with synthesis for an increasing
number of processes using classical, non-parameterized methods. 

 We have considered the problem in detail for token-passing systems, including token rings.
 Using results
 from parameterized verification, we showed that the parameterized
 synthesis problem reduces to distributed synthesis in a small network
 of isomorphic processes with fairness constraints on token
 passing. Unfortunately, the
 distributed synthesis problem remains undecidable, even for small
 token rings.


Regardless of this negative result, we managed to synthesize an
actual --- albeit very small --- example of a parameterized arbiter.  To
this end, we used Schewe and Finkbeiner's results on bounded
synthesis.  In theory, this approach will eventually find
an implementation if it exists. In practice, this currently only works for small 
implementations. One line of future work will be on making synthesis feasible
for larger systems --- together with Khalimov, we recently started research in that
direction~\cite{Khalimov13}, and managed to reduce synthesis time by
several orders of magnitude by using modularity and abstraction
techniques. We plan to extend research into more efficient encoding
techniques, as well as the integration of ideas from the lazy synthesis 
approach~\cite{Finkbeiner12}.

For unrealizable specifications, our approach will run forever.
It is an interesting question whether it could be combined with incomplete
methods to check unrealizability. 

We note that the topologies we considered limit communication
between processes, and therefore also the possible solutions.  For
our running example, processes give grants only when they hold the
token. In a token ring, this 
means that response time increases linearly with the number of
processes, something that can be avoided in other topologies. We can widen the class of topologies that we can synthesize by using more general results on parameterized verification.

To this end, we have given an incomplete
overview of results for parameterized model checking, and ideas for
how to lift these results to parameterized synthesis. In particular,
the approach presented in this article can be seen as a framework for
lifting cutoff-based reduction techniques from parameterized
verification to parameterized synthesis.

\section*{Acknowledgments.} Many thanks to Leonardo de Moura for his
help with little known features of Z3.  We thank the members of
ARiSE, particularly Igor Konnov, Sasha Rubin, Helmut Veith, and Josef Widder, 
for stimulating discussions on parameterized verification and synthesis, 
and Bernd Finkbeiner for discussions on distributed and
bounded synthesis. Finally, thanks to Hossein Hojjat, Ayrat Khalimov, and the anonymous referees for useful comments on 
drafts of this paper.

\bibliographystyle{alpha}
\bibliography{references,local}

\newcommand{\etalchar}[1]{$^{#1}$}
\begin{thebibliography}{BGJ{\etalchar{+}}07b}

\bibitem[Abd12]{Abdulla12}
P.~A. Abdulla.
\newblock Regular model checking.
\newblock {\em STTT}, 14(2):109--118, 2012.

\bibitem[AE98]{Attie98}
P.~C. Attie and E.~A. Emerson.
\newblock Synthesis of concurrent systems with many similar processes.
\newblock {\em ACM Trans. Program. Lang. Syst.}, 20(1):51--115, January 1998.

\bibitem[AJKR14]{AJKR14}
B.~Aminof, S.~Jacobs, A.~Khalimov, and S.~Rubin.
\newblock Parameterized model checking of token-passing systems.
\newblock In {\em VMCAI}, volume 8318 of {\em LNCS}, pages 262--281. Springer,
  2014.

\bibitem[AK86]{Apt86}
K.~Apt and D.~Kozen.
\newblock Limits for automatic verification of finite-state concurrent systems.
\newblock {\em Inf. Process. Lett.}, 22:307--309, 1986.

\bibitem[BBF{\etalchar{+}}12]{Bohy12}
A.~Bohy, V.~Bruyère, E.~Filiot, N.~Jin, and J.-F. Raskin.
\newblock Acacia+, a tool for {LTL} synthesis.
\newblock In {\em CAV}, volume 7358 of {\em LNCS}, pages 652--657. Springer,
  2012.

\bibitem[BGJ{\etalchar{+}}07a]{Bloem07b}
R.~Bloem, S.~Galler, B.~Jobstmann, N.~Piterman, A.~Pnueli, and M.~Weiglhofer.
\newblock Automatic hardware synthesis from specifications: A case study.
\newblock In {\em DATE}, pages 1188--1193, 2007.

\bibitem[BGJ{\etalchar{+}}07b]{Bloem07}
R.~Bloem, S.~Galler, B.~Jobstmann, N.~Piterman, A.~Pnueli, and M.~Weiglhofer.
\newblock Specify, compile, run: Hardware form {PSL}.
\newblock {\em ENTCS}, 190(4):3--16, 2007.

\bibitem[BHV08]{Bouajjani08}
A.~Bouajjani, P.~Habermehl, and T.~Vojnar.
\newblock Verification of parametric concurrent systems with prioritised {FIFO}
  resource management.
\newblock {\em Formal Methods in System Design}, 32(2):129--172, 2008.

\bibitem[BJNT00]{Bouajjani00}
A.~Bouajjani, B.~Jonsson, M.~Nilsson, and T.~Touili.
\newblock Regular model checking.
\newblock In {\em CAV}, volume 1855 of {\em LNCS}, pages 403--418. Springer,
  2000.

\bibitem[CGJ97]{Clarke97}
E.~M. Clarke, O.~Grumberg, and S.~Jha.
\newblock Verifying parameterized networks.
\newblock {\em ACM Trans. Program. Lang. Syst.}, 19(5):726--750, 1997.

\bibitem[Chu62]{Church62}
A.~Church.
\newblock Logic, arithmetic and automata.
\newblock In {\em Proceedings International Mathematical Congress}, 1962.

\bibitem[CTTV04]{Clarke04c}
E.~M. Clarke, M.~Talupur, T.~Touili, and H.~Veith.
\newblock Verification by network decomposition.
\newblock In {\em CONCUR}, volume 3170 of {\em LNCS}, pages 276--291. Springer,
  2004.

\bibitem[CTV06]{Clarke06}
E.~M. Clarke, M.~Talupur, and H.~Veith.
\newblock Environment abstraction for parameterized verification.
\newblock In {\em VMCAI}, volume 3855 of {\em LNCS}, pages 126--141. Springer,
  2006.

\bibitem[CTV08]{Clarke08}
E.~M. Clarke, M.~Talapur, and H.~Veith.
\newblock Proving ptolemy right: The environment abstraction framework for
  model checking concurrent systems.
\newblock In {\em TACAS}, volume 4963 of {\em LNCS}, pages 33--47. Springer,
  2008.

\bibitem[DMB08]{Moura08}
L.~De~Moura and N.~Bj{\o}rner.
\newblock Z3: An efficient {SMT} solver.
\newblock In {\em TACAS}, volume 4963 of {\em LNCS}, pages 337--340. Springer,
  2008.

\bibitem[EK00]{Emerson00}
E.~A. Emerson and V.~Kahlon.
\newblock Reducing model checking of the many to the few.
\newblock In {\em CADE}, volume 1831 of {\em LNCS}, pages 236--254. Springer,
  2000.

\bibitem[EK03]{Emerson03}
E.~A. Emerson and V.~Kahlon.
\newblock Exact and efficient verification of parameterized cache coherence
  protocols.
\newblock In {\em CHARME}, volume 2860 of {\em LNCS}, pages 247--262. Springer,
  2003.

\bibitem[EK04]{Emerson04}
E.~A. Emerson and V.~Kahlon.
\newblock Parameterized model checking of ring-based message passing systems.
\newblock In {\em CSL}, volume 3210 of {\em LNCS}, pages 325--339. Springer,
  2004.

\bibitem[EN03]{Emerso03}
E.~A. Emerson and K.~S. Namjoshi.
\newblock On reasoning about rings.
\newblock {\em Foundations of Computer Science}, 14:527--549, 2003.

\bibitem[ES90]{Emerson90}
E.~A. Emerson and J.~Srinivasan.
\newblock A decidable temporal logic to reason about many processes.
\newblock In {\em PODC}, pages 233--246, New York, NY, USA, 1990. ACM.

\bibitem[FJ12]{Finkbeiner12}
B.~Finkbeiner and S.~Jacobs.
\newblock Lazy synthesis.
\newblock In {\em VMCAI}, volume 7148 of {\em LNCS}, pages 219--234. Springer,
  2012.

\bibitem[FS05]{Finkbeiner05}
B.~Finkbeiner and S.~Schewe.
\newblock Uniform distributed synthesis.
\newblock In {\em LICS}, pages 321--330. IEEE Computer Society Press, 2005.

\bibitem[FS13]{Finkbeiner12a}
B.~Finkbeiner and S.~Schewe.
\newblock Bounded synthesis.
\newblock {\em STTT}, 15(5-6):519--539, 2013.

\bibitem[GS92]{German92}
S.~M. German and A.~P. Sistla.
\newblock Reasoning about systems with many processes.
\newblock {\em J. ACM}, 39(3):675--735, 1992.

\bibitem[HBR09]{Hanna09}
Y.~Hanna, S.~Basu, and H.~Rajan.
\newblock Behavioral automata composition for automatic topology independent
  verification of parameterized systems.
\newblock In {\em ESEC/SIGSOFT FSE}, pages 325--334. ACM, 2009.

\bibitem[JB12]{JacobsBloem12}
S.~Jacobs and R.~Bloem.
\newblock Parameterized synthesis.
\newblock In {\em TACAS}, volume 7214 of {\em LNCS}, pages 362--376. Springer,
  2012.

\bibitem[KIG05]{Kahlon05}
V.~Kahlon, F.~Ivancic, and A.~Gupta.
\newblock Reasoning about threads communicating via locks.
\newblock In {\em CAV}, volume 3576 of {\em LNCS}, pages 505--518. Springer,
  2005.

\bibitem[KJB13a]{Khalimov13a}
A.~Khalimov, S.~Jacobs, and R.~Bloem.
\newblock {PARTY} parameterized synthesis of token rings.
\newblock In {\em CAV}, volume 8044 of {\em LNCS}, pages 928--933. Springer,
  2013.

\bibitem[KJB13b]{Khalimov13}
A.~Khalimov, S.~Jacobs, and R.~Bloem.
\newblock Towards efficient parameterized synthesis.
\newblock In {\em VMCAI}, volume 7737 of {\em LNCS}, pages 108--127. Springer,
  2013.

\bibitem[KKW10]{Kaiser10}
A.~Kaiser, D.~Kroening, and T.~Wahl.
\newblock Dynamic cutoff detection in parameterized concurrent programs.
\newblock In {\em CAV}, volume 6174 of {\em LNCS}, pages 645--659. Springer,
  2010.

\bibitem[KM95]{Kurshan95}
R.~P. Kurshan and K.~L. McMillan.
\newblock A structural induction theorem for processes.
\newblock {\em Information and Computation}, 117(1):1--11, 1995.

\bibitem[KP09]{Katz09}
G.~Katz and D.~Peled.
\newblock Synthesizing solutions to the leader election problem using model
  checking and genetic programming.
\newblock In {\em HVC}, volume 6405 of {\em LNCS}, pages 117--132. Springer,
  2009.

\bibitem[KV05]{KupfermanV05}
O.~Kupferman and M.~Y. Vardi.
\newblock Safraless decision procedures.
\newblock In {\em FOCS}, pages 531--542. IEEE Computer Society, 2005.

\bibitem[PPS06]{Piterm06b}
N.~Piterman, A.~Pnueli, and Y.~Sa{\'{}}ar.
\newblock Synthesis of reactive(1) designs.
\newblock In {\em VMCAI}, volume 3855 of {\em LNCS}, pages 364--380. Springer,
  2006.

\bibitem[PR89]{Pnueli89}
A.~Pnueli and R.~Rosner.
\newblock On the synthesis of a reactive module.
\newblock In {\em POPL}, pages 179--190. ACM Press, 1989.

\bibitem[PR90]{Pnueli90}
A.~Pnueli and R.~Rosner.
\newblock Distributed systems are hard to synthesize.
\newblock In {\em FOCS}, pages 746--757. IEEE Computer Society, 1990.

\bibitem[PRZ01]{Pnueli01}
A.~Pnueli, S.~Ruah, and L.~Zuck.
\newblock Automatic deductive verification with invisible invariants.
\newblock In {\em TACAS}, volume 2031 of {\em LNCS}, pages 82--97. Springer,
  2001.

\bibitem[PXZ02]{Pnueli02}
A.~Pnueli, J.~Xu, and L.~Zuck.
\newblock Liveness with (0,1,$\infty$)- counter abstraction.
\newblock In {\em CAV}, volume 2404 of {\em LNCS}, pages 107--122. Springer,
  2002.

\bibitem[Sch14]{Schewe2013}
S.~Schewe.
\newblock Distributed synthesis is simply undecidable.
\newblock {\em Inf. Process. Lett.}, 114(4):203--207, 2014.

\bibitem[SF06]{Schewe06}
S.~Schewe and B.~Finkbeiner.
\newblock Synthesis of asynchronous systems.
\newblock In {\em LOPSTR}, volume 4407 of {\em LNCS}, pages 127--142. Springer,
  2006.

\bibitem[Suz88]{Suzuki88}
I.~Suzuki.
\newblock Proving properties of a ring of finite state machines.
\newblock {\em Inf. Process. Lett.}, 28(4):213--214, 1988.

\bibitem[Voj07]{Vojnar07}
T.~Vojnar.
\newblock Cut-offs and automata in formal verification of infinite-state
  systems, 2007.
\newblock Habilitation thesis.

\bibitem[ZP04]{Zuck04}
L.~D. Zuck and A.~Pnueli.
\newblock Model checking and abstraction to the aid of parameterized systems (a
  survey).
\newblock {\em Computer Languages, Systems {\&} Structures}, 30(3-4):139--169,
  2004.

\end{thebibliography}
\vspace{-32 pt}
\end{document}